\documentclass[superscriptaddress, amsmath,amssymb, pre, twocolumn]{revtex4-1}
\usepackage{graphicx}
\usepackage{dcolumn}
\usepackage{bm}
\usepackage{hyperref}
\usepackage{color}
\usepackage{amsmath}
\usepackage{amssymb}
\usepackage{tikz}

\usepackage{ulem}

\DeclareUnicodeCharacter{2009}{\,}

\begin{document}

\title{Emergent conformational properties of end-tailored transversely propelling polymers}
\author{K.~R.~Prathyusha}
\email{krprathyusha@gmail.com}
\affiliation{Max Planck Institute for Dynamics and Self-Organization, D-37077 Göttingen, Germany}%
\affiliation{Center for Softmatter Physics and its Applications, University of Beihang, Beijing, China}

\author{Falko Ziebert}
\email{f.ziebert@thphys.uni-heidelberg.de}
\affiliation{Institute for Theoretical Physics, Heidelberg University, D-69120 Heidelberg, Germany}

\author{Ramin Golestanian}
\email{ramin.golestanian@ds.mpg.de}
\affiliation{Max Planck Institute for Dynamics and Self-Organization, D-37077 Göttingen, Germany}
\affiliation{Rudolf Peierls Centre for Theoretical Physics, University of Oxford, Oxford OX1 3PU, United Kingdom}

\date{\today}

\begin{abstract} 
We study the dynamics and conformations of a single active semiflexible polymer 
whose monomers experience a propulsion force perpendicular to the local tangent, 
with the end beads being different from the inner beads (``end-tailored''). 
Using Langevin simulations, we demonstrate that, apart from sideways motion, 
the relative propulsion strength between the end beads and the polymer backbone 
significantly changes the conformational properties of the polymers as a function
of bending stiffness, end-tailoring and propulsion force.
Expectedly, for slower ends the polymer curves away from the moving direction,
while faster ends lead to opposite curving, in both cases slightly 
reducing the center of mass velocity compared to a straight fiber.
Interestingly, for faster end beads there is a rich and dynamic morphology diagram:
the polymer ends may get folded together to 2D loops or
hairpin-like conformations that rotate due to their asymmetry in shape
and periodic flapping motion around a rather straight state during full propulsion is also possible. 
We rationalize the simulations using scaling and kinematic arguments 
and present the state diagram of the conformations.  
Sideways propelled fibers comprise a rather unexplored and versatile class of self-propellers,
and their study will open novel ways for designing, e.g. motile actuators or
mixers in soft robotics.
\end{abstract}


\maketitle
\section{Introduction}

Self-propelled particles are the most studied realisation of active matter \cite{roadmap},
i.e.~systems in out-of-equilibrium states using energy sources or fluxes 
to perform tasks, in this case to move in an autonomous and persistent fashion.
Prominent examples from the living world are swimming ciliated bacteria \cite{Berg73,Lauga2009}
gliding myxobacteria \cite{myxoglide,peruanibaer} and
cytoskeletal filaments transported on motor 
carpets~\cite{howardMTglide,vschaller-10}
In the last 15 years a variety of artificial self-propellers 
has been developed, mostly microswimmers, with elegant examples being
catalytic nanorods~\cite{active_Pt_Aurods} and phoretically 
or demixing-driven colloidal Janus particles \cite{Ebbens-Ramin-PRE-12,HowseRamin,Soto-Ramin-PRL-14,BechingerJPCM}.

Most self-propellers studied so far are either spherical or the propulsion direction is 
associated with the major body axis. 
However, recently also sideways propelled objects have been developed.
In the colloidal realm, examples are sideways propelling short ellipsoids \cite{pietro-sagues-small10}
and cylinders \cite{vutukuri-huck-collective-transverse-rods-SM16}
as well as straight and bent Janus micro-rods \cite{vrao-clasen-jphys-D-19}
and 3D printed helical or more complex shapes 
\cite{3d_printinghelix}.
Active Janus beads were recently assembled to flexible chains 
in an electric field \cite{vrao-huck-scirepD-17}, but
so far only in staggered orientation with no sideways propulsion.
Other examples are anisotropic rollers, for instance flexible 
fibers propelled by a flux of heat or swelling agent to roll along their long axis
\cite{falko-nature-material-18,Falko_fiberiod-2020} 
as well as light-driven rolling helices and spring actuators \cite{rolling-spring-actuator-ChemMat-21}.   

While most active colloidal systems are mechanically rigid so far, 
it is evident that a certain deformability of active objects will lead to more complex behavior 
with possible applications in soft robotics \cite{granick-janus-chain-nat-mat16}. 
Minimal models of active polymers, composed of stresslets or propelled beads,  
were shown to exhibit translational or rotational motion or spiraling 
\cite{gayathri-12,riseleholder-15,Winkler-conformational-property-16,hiang-14a}. 

While tangentially propelled polymers typically deform only weakly, 
as the main force is along the backbone,
it is clear that for transversely propelled slender objects, deformations can easily be excited
-- similar to but different from fibers sedimenting or actively dragged through a viscous environment \cite{Ignacio_filamentdeformation_PRL_05,
Ali_nadim_deformatin_orientation_phys_fluids_94,
Netz_EPL_05,filamentgravity_exp_marchettiPRF_19} --
and in turn will interfere with the propulsion. 
Motivated by this reasoning 
we here study the following principal questions: What is the interplay of active forces
and flexibility in a sideways propelled slender object? And can one tailor the properties
to dynamically obtain certain shapes or even complex behavior that may be useful to perform certain tasks? 
To answer these questions we propose and study a simple coarse-grained active polymer model
with two main ingredients: first, the (local) propulsion force is pointing in the direction 
normal to the (local) backbone direction. And second, we consider what we call ``end-tailoring'',
i.e.~the end beads are made different from the backbone beads, which
upon self-propulsion will induce an initial bending deformation and, depending on parameters and design,
complex behavior.

The work is organized as follows. 
Sec.~II introduces 
the coarse-grained model 
for an end-tailored semi-flexible filament subject to active forces  perpendicular to its contour. 
In Sec.~III, we briefly discuss the propulsion behavior without end-tailoring.
Sec.~IV studies the case of slower end beads where we extensively analyze the 
propulsion induced bending and its feedback on velocity.
Sec.~V studies the case of faster end beads, where the behaviour turns out to be much richer,
leading to closed shapes and even periodic flapping during propulsion.
Finally, in Sec.~VI we summarize our main findings and comment on future directions.

\section{Model and computational details}

\begin{figure}
	\includegraphics[width=0.45\textwidth]{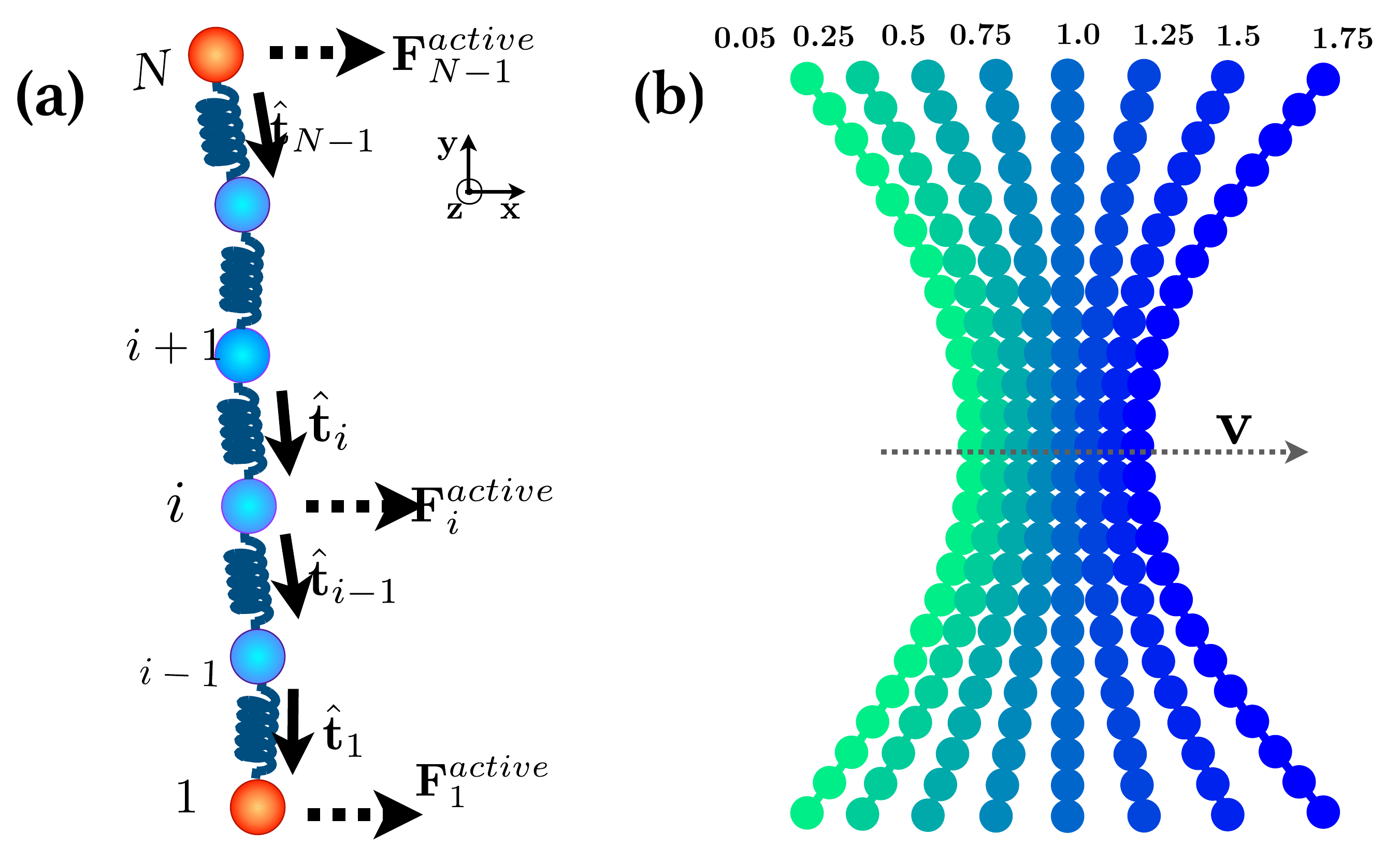}
	\caption{ 
	(a) Schematic of the model for a transversely propelling end-tailored  filament. 
	Briefly, bead $i$ gets sideways active force contributions from  
	the tangents ${\bf t}_i$ and ${\bf t}_{i-1}$,
	whereas the end beads $1$ and $N$ get the propulsion force only from the tangents  
	${\bf t}_1$ and ${\bf t}_{N-1}$, respectively. More importantly, 
	while all inner beads (blue) have the same absolute value of propulsion 
	force $|{\bf F}_i^{active}|=f_p$, 
	the end beads (red) experience a smaller (parameter $\alpha<1$)
	or larger (parameter $\alpha>1$) propulsion force $\alpha f_p$.  
	(b) Conformations of steadily propelling polymers for various  values of $\alpha$. 
	For $\alpha =1$ the polymer is straight (in the absence of noise), 
	end-tailored propulsion induces curving. 
	The propulsion direction is indicated by the arrow and $\mathbf{v}$.
	Parameters: bending stiffness $\kappa=2$, $f_p=2$ and $\alpha$ as indicated. 
	}
	\label{fig:schematic-alpha-conformation}
\end{figure}

We model a two-dimensional (2D) semiflexible polymer, consisting of $N$ beads 
of diameter $\sigma$ and mass $m_i=m$ with coordinates ${\bf r}_i$ $(i =1, . . . , N)$. 
The advantage of employing a polymeric model is that one can easily tune both 
the aspect ratio (via the length) and the stiffness of the object. 
An active propulsion force that is 
perpendicular to the local tangent along the polymer contour is implemented as explained below.  
In the stiff limit, the model corresponds to a transversely propelled rod, 
but when the polymer curves by the prevalent forces, complex behaviour 
is anticipated from the interplay of propulsion and shape. 
A schematic of the model is given in Fig.~\ref{fig:schematic-alpha-conformation}a. 

The time evolution of a monomer is given by the following Langevin equation, 
 \begin{equation}
 m_i\frac{d{\bf v}_i}{dt}={\bf F}^{poly}_i+{\bf F}_i^{fric}+{\bf F}_i^{noise}+{\bf F}_i^{active},
 \label{eq:eqofmotion}
 \end{equation}
with  ${\bf v}_i=\frac{d{\bf r}_i}{dt}$ the velocity of bead $i$ and the forces on the right hand side
being a conservative force modeling the polymeric structure, a friction force, 
a stochastic force and the active propulsion force. 

The conservative force 
${\bf F}^{poly}_i= -\frac{\partial U(r)}{\partial r_i}{ \mathbf {\hat r}_{ij}}$, 
where $r=\left|\mathbf{r}_{ij}\right|=\left|\mathbf{r}_i-\mathbf{r}_j\right|$
is the distance between two beads (and ${\mathbf {\hat r}}_{ij}= \mathbf{r}_{ij}/r$),
follows the potential $U(r)=U_{WCA}+U_{s}+U_{b}$. 
Excluded volume interactions between the  beads are
modeled  via the Weeks-Chandler-Anderson potential \cite{weeks-chandler-71}, 
\begin{equation}
U_{WCA}=4\epsilon\left[\left(\frac{\sigma}{r}\right)^{12}-\left(\frac{\sigma}{r}\right)^{6}+\frac{1}{4}\right],
\end{equation}
which vanishes  beyond distances greater than $2^{1/6}\sigma$.  
Here $\epsilon$ measures the strength of the steric repulsion, $\sigma$ is the size of the bead, 
and both are set to $1$ in the following. 
This scaling of energy in terms of $\epsilon$ and length in terms of $\sigma$
implies the reduced unit of time $\tau=\sqrt{\frac{m\sigma^2}{\epsilon}}$. 

Every connected bead in a polymer interacts via a two-body harmonic potential \cite{kkremer-90}, 
\begin{equation}
U_{s}=k_b \big(r- R_0\big)^2.
\label{eq:bond}
\end{equation}
Here $R_0=\sigma$ is the maximum bond length and $k_b=4000\epsilon/\sigma^2$ 
is the bond stiffness (this large value making the chain effectively inextensible). 
These parameters ensure the bond length to be $b\simeq\sigma$ and 
the polymer contour length to be 
$ L\simeq(N-1)\sigma $ to very good approximation.
The three-body bending potential $U_b$ is employed  for every consecutive triplet of monomers $i$, $i+1$, $i+2$, 
 \begin{equation}
 U_{b}= \kappa \big(\theta -\theta_0 \big)^2.
 \end{equation}
It measures the energy cost for departures of the angle $\theta$   
between consecutive bond vectors, 
$\theta= \arccos({\bf \hat r}_{i,i+1}\cdot{\bf \hat r}_{i+2,i+1})$,
from its equilibrium value $\theta_0 =\pi$. Here $\kappa$ is the bending stiffness,
related  to the continuum bending stiffness, $\tilde \kappa$, via $\kappa = \tilde \kappa /2\sigma$. 

${\bf F}_i^{fric} = -\xi{\bf v}_i$ accounts for the frictional force 
with  coefficient $\xi$, which is set to $1$ unless otherwise specified. 
Note that for simplicity we use an isotropic friction here, but it would be straightforward
to implement anisotropic friction (e.g.~for hydrodynamic systems). 
The stochastic (in principle not necessarily thermal) force, ${\bf F}^{noise}_i$, 
is modelled as a white noise with zero mean and variance  $ 4 k_BT \xi/\Delta t $. 
In most of our simulation the strength of the noise is zero (deterministic limit), 
unless otherwise specified. 
 
Finally, we implement an active propulsion force 
as follows : 
all  the beads of the backbone 
$(i =2, . . . , N-1)$ 
experience the force 
\begin{equation}
{\bf F}_i^{active}=  \hat{ \bf z} \times \frac{\hat{\bf t}_i +\hat{\bf t}_{i -1}}{2} {f_p} ,
\label{activeforceb}
\end{equation}
Here $\hat {\bf t}_i= \frac{{\bf r}_i-{\bf r}_{i+1}}{|{\bf r}_i-{\bf r}_{i+1}|}$ is the 
local tangent between beads $i$ and $i+1$. 
Hence both neighbours of bead $i$ contribute equally, and the respective tangents
are averaged. $\hat{ \bf z}$ is the unit vector normal to the 2D plane of motion,
the cross product defining the normal vector to the contour.
$f_p$ is the propulsion strength.  
The active force defined by Eq.~\ref{activeforceb} is hence distributed equally between 
the beads and acts perpendicular to the contour of the polymer. 

As the end beads  $i=1$ and $i=N$ have only one nearest neighbour, 
we define their active forces as 
 \begin{equation}
{\bf F}_{1}^{active}=  \hat{ \bf z} \times \hat{\bf t}_{1}  \alpha f_{p}\;\;\,\,,\,\, \;\;
 {\bf F}_{N}^{active}=  \hat{ \bf z} \times \hat{\bf t}_{N-1} \alpha f_{p},
\label{activeforce-end}
\end{equation}
where in addition, we introduced the parameter $\alpha$. 
This parameter implements the end-tailoring of 
the polymer, i.e.~differences in the propulsion and/or material properties of the ends,
and will play a key role  in the emergence of complex shapes during propulsion.
Clearly, $\alpha = 1$ corresponds to homogeneous propulsion,
where one expects -- in the absence of noise and after having attained a straight conformation
from a given initial condition -- a center of mass (c.o.m.) velocity of $V=v_{cm}=f_p/\xi$.
In case $\alpha<1$ ($>1$), the end beads move slower (faster), inducing curving while propelling 
as shown in Fig.~\ref{fig:schematic-alpha-conformation}b. For higher propulsion strengths 
more complex conformational dynamics emerges as investigated in the following.

The effect of a different propulsion of the ends, 
modelled here by the parameter $\alpha$, 
should always be present in a real system, 
since the ends are typically slightly different from the bulk. 
However, this is not very generic
and typically a weak and subtle effect.
To experimentally realize large deviations from $\alpha=1$, it is best to tailor
the ends specifically. 
Consider a chain of Janus colloids, similar
as developed in \cite{vrao-huck-scirepD-17}, 
but not with a staggered but with a polarized orientation of the beads.
Then using colloids with  less catalytic activity at the ends 
will induce a well-defined end-tailoring $\alpha<1$.
Conversely, $\alpha>1$ can be achieved for end colloids having larger catalytic activty. 
In an even simpler fashion, the case $\alpha\simeq0$ can be achieved 
by putting passive loads at the ends.

Concerning  parameter ranges,
in the following we vary the number of beads in the polymer between $N \in \{5, 10, 25, 50\}$, 
the filament stiffness $\kappa = 1-500$, 
the magnitude of the active force $f_p=0.01-30$ 
and $\alpha=0.05-2$.  
The parameters can be combined to a dimensionless number,  
$B=\frac{f_p L^3}{\sigma^2\kappa}$ (note that $\sigma=1$), 
which characterizes the ratio of (bulk) propulsion to bending rigidity. For all  numerical results shown in the following, the equation of motion, Eq.~\ref{eq:eqofmotion}, 
was integrated using LAMMPS~\cite{splimpton-95},
with time step $\Delta t = 10^{-4} \tau$.

\section{Homogeneous transversely propelled fiber ($\alpha=1$)}

\begin{figure}
	\includegraphics[width=0.5\textwidth]{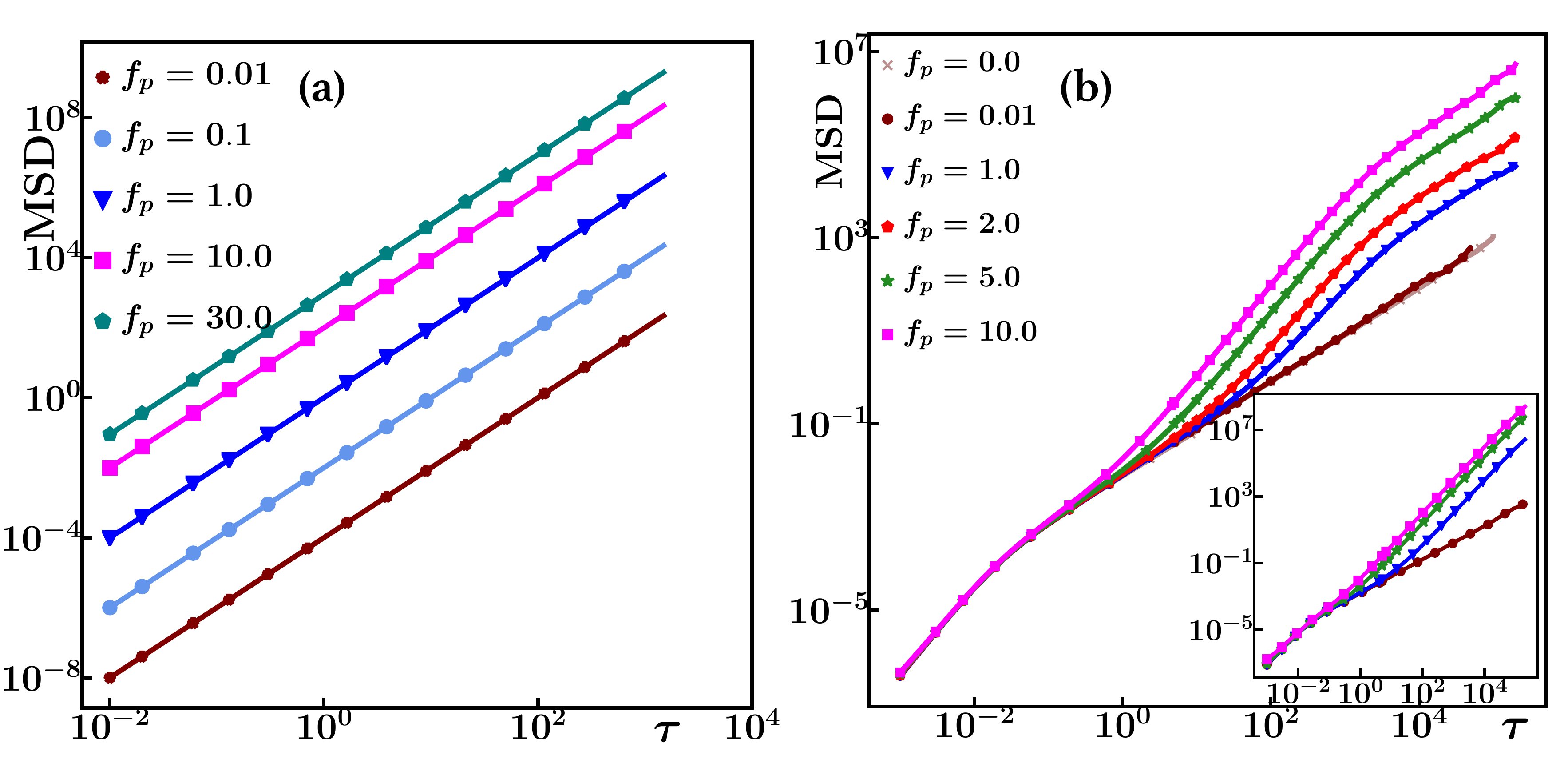}
	\caption{ The mean squared displacements ($\rm {MSD}$) of transversely propelled polymers 
	for different propulsion strengths.
	(a) Perfect self-propulsion without noise, 	the solid line  representing 
	$(f_p \tau/\xi)^2$. Parameters: $N=25$,  $\kappa=100$, $\xi= 1$. 
	(b) The case with noise present. The MSD displays up to four regimes; see main text. 
	Parameters: $N=5$, $\kappa=100$, $\xi=100$, $k_BT=1$.
	The inset is for a longer polymer, $N=25, \kappa=500$.
	}
	\label{fig:alpha-2-msd-vcm}
\end{figure}

We first study the case of a homogeneously propelled fiber, i.e. $\alpha=1$. 
Directed vs.~diffusive motion can be quantified 
by the  mean squared displacement 
of the center of mass of the filament, 
$\rm {MSD}=\Big\langle \Big ({\bf r}_{cm}(t+\tau)-{\bf r}_{cm}(t)\Big )^2\Big \rangle$, as a 
function of time. 
In the absence of  noise, one expects all monomers to move with the same propulsion force, 
leading to a translational motion perpendicular to the long body axis 
with center of mass (c.o.m.)  velocity ${ \bf v}_{cm}=f_p/\xi$. %
Such a ballistic dependence on time 
proportional to the strength of the propulsion force
is a hallmark of active propulsion
and is clearly visible in Fig.~\ref{fig:alpha-2-msd-vcm}a
in the absence of noise.
Introducing noise changes only weakly the conformation of a stiff polymer 
(see also the next section), but it induces long-time orientational decorrelation 
of the whole object. This is shown in
Fig.~\ref{fig:alpha-2-msd-vcm}b for stiff sideways propelled filaments 
in the presence of noise ($k_BT=1$).

Note that this behavior is 
similar to what is observed for tangentially propelled objects (or active point particles).
As Fig.~\ref{fig:alpha-2-msd-vcm}b exemplifies for short filaments $(N=5)$ 
and high friction $(\xi=100)$,
 the MSD exhibits 
four  power-law regimes for sufficiently high propulsion strength:  
a short-time inertial regime, an intermediate diffusive, an active ballistic and 
a long-time diffusive regime. These regimes are  well understood within 
the Langevin theory of active Brownian motion \cite{HowseRamin,Ramin_PRL_anomalous_diffusion-09,inertia_natcom_Lowen-2018}. 
The cross-over time between the initial inertial ballistic and  
the primary diffusive behaviour is $\sim m/\xi $,  
independent of propulsion force. 
The extent of the short-time diffusive regime depends on the active force as the  
diffusive to active ballistic transition occurs at  $\sim D/(2V^2)$~\cite{inertia_natcom_Lowen-2018},
with $D=\frac{k_BT}{\xi}$ the diffusion coefficient and $V$ the c.o.m.~velocity. 
Directed motion dominates in the third regime and the MSD yields 
a superdiffusive behaviour (approximately ballistic, ${\rm MSD} \propto t^2 $). 
Finally, a long time  diffusive regime is attained, due to the 
decorrelation of the orientation, with an enhanced diffusion coefficient.  
This occurs at time $\sim D_r^{-1}$~\cite{HowseRamin,Ramin_PRL_anomalous_diffusion-09,inertia_natcom_Lowen-2018}
(with $D_r$ the rotational diffusion coefficient)
being the time scale at which the object 
loses its memory of the initial orientation. For a stiff polymer this time 
rapidly increases with length, $\sim$  $\frac{ \xi L^3}{k_BT}$ \cite{doi-edwards-86}.  
In fact, the inset of  Fig.~\ref{fig:alpha-2-msd-vcm}b shows the case of $N=25$,
where the long time diffusive regime is difficult to attain in our simulations.  

From Fig.~\ref{fig:alpha-2-msd-vcm}b, it is clear that the propulsion force
has to be substantial to see the intermediate, self-propelled regime.
While the interplay of weak noise-induced bending and propulsion may be interesting,
we focus here on the opposite limit, i.e.~strong propulsion where the noise
is rather irrelevant. Except if otherwise stated, the rest of the paper focuses
on the deterministic dynamics without noise.

\section{Slower end beads  ($\alpha<1$) }

\begin{figure}
\includegraphics[width=0.5\textwidth]{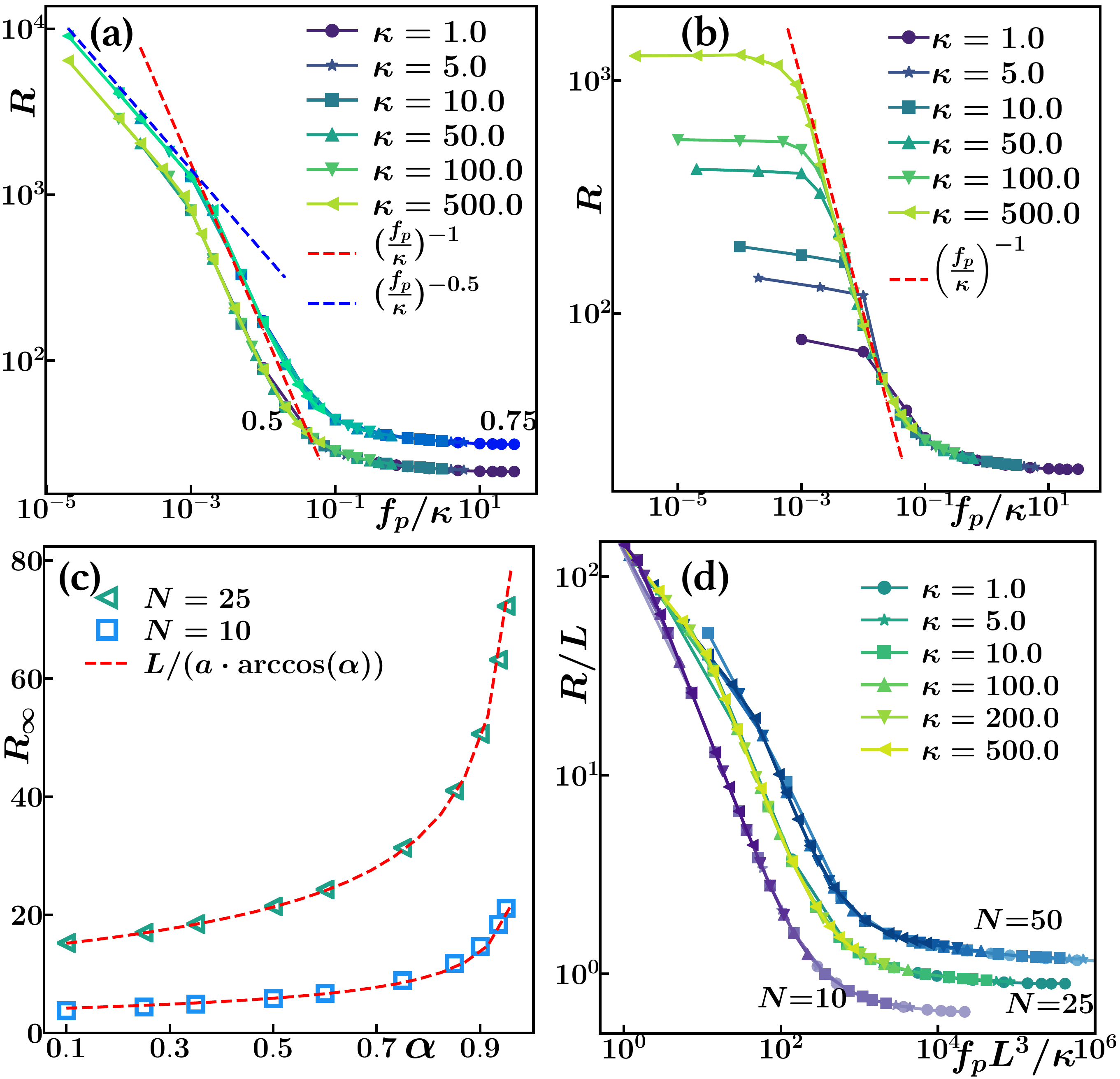}
\caption { 
Curvature effects for $\alpha <1$.
(a) Radius of curvature $R$ as a function of $f_p/\kappa$ 
in the absence of noise, for length $N=25$ and two different values of $\alpha$.  
For a given $\alpha$, the curves collapse with two different scaling regimes 
and a saturation regime.  
(b) In the presence of noise ($k_BT=0.01$) the first scaling regime is absent 
and replaced by a fluctuation-induced small-curvature plateau 
(length $N=25$, $\alpha =0.5$). 
(c) The high-driving force saturation value of the radius of curvature ($R_\infty$, cf.~panel a)
as a function of end-tailoring parameter $\alpha$ 
showing the behavior  $R_\infty\propto1/\arccos(\alpha)$, see text.  
(d) Rescaled radius of curvature $R/L$  as a function of the dimensionless 
driving force $B=f_pL^3/\kappa$, for $\alpha=0.5$ and different lengths $N$. 
}
\label{fig:R_of_fp}
\end{figure}

\begin{figure*}
\includegraphics[width=0.9\textwidth]{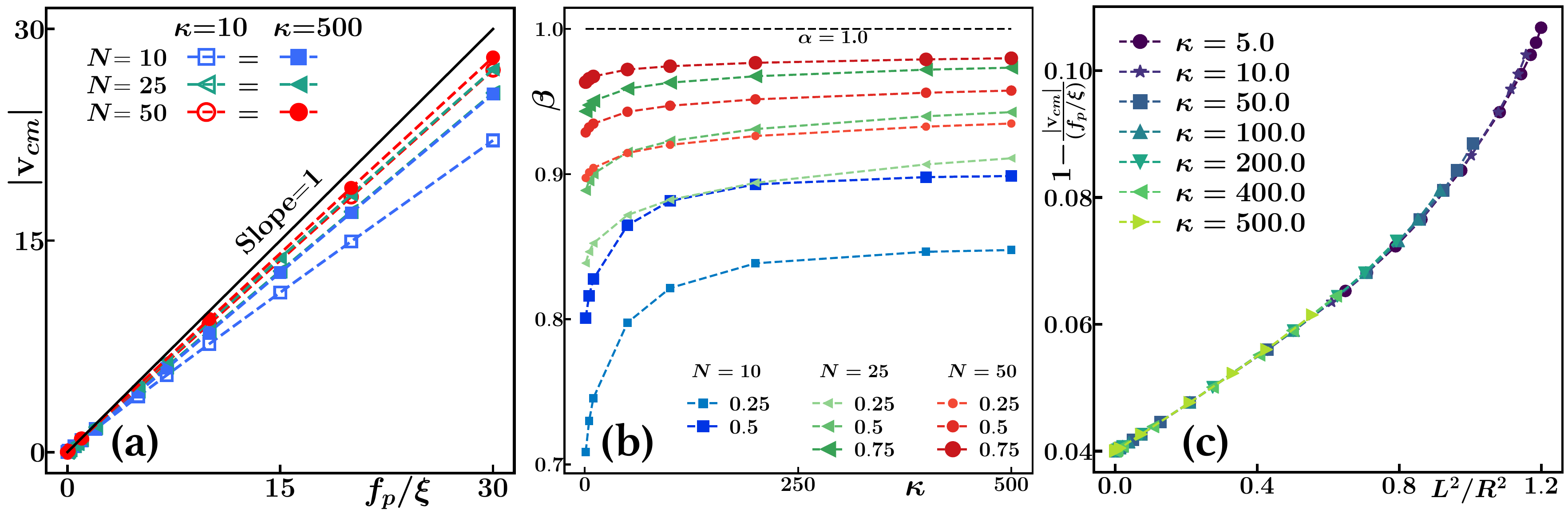} 
\caption{
Feedback of curvature on velocity for $\alpha<1$.
(a) $|v_{cm}|$ as a function of $f_p$
for $\alpha=0.5$, 
two stiffness values, $\kappa=10$ (open symbols),
 $\kappa=500$ (filled symbols), 
and for 
 $N=10$~(blue squares),   $N=25$~(green triangles),   $N=50$~(red circles).
 The solid black line of unit slope corresponds to  the rigid limit $|v_{cm}|=\frac{f_{p}}{\xi}$. 
 Slopes $\neq0$ define the velocity reduction factor $\beta$.
(b) $\beta$ vs.~filament stiffness $\kappa$, 
for different lengths ($N=10$~$(\blacksquare)$,~$N=25~( \bullet)$,~$N=50~(\blacktriangledown )$) 
and different end-tailoring parameters $\alpha$ (see legend).
(c) Approximate scaling for the velocity reduction factor $\beta$, see text.}
\label{fig:alpha-lessthan2-msd-beta}
\end{figure*}

In the following we mainly focus on the noise-less limit to carve out the features 
of the end-induced curvature effects. We first study the case when $\alpha  < 1 $,
i.e.~when the end beads experience a smaller propulsion force than the polymer's backbone. 
As for the case $\alpha=1$, there is propulsion in the direction perpendicular 
to the backbone, but in addition
the distribution of the active forces produces a curving along the propulsion direction, 
as shown in Fig.~\ref{fig:schematic-alpha-conformation}b:
since the end beads move more slowly, they lag behind the backbone,  
and the fiber starts to bend (see also Suppl.~movie 1).
The conformation of the polymer transiently evolves with time
and rapidly attains a stable bent conformation
from the balance of the active and the elastic forces, 
while still propelling transversely with constant velocity.

For not too long polymers ($N\lesssim25$) the steady conformation is close to a circular arc. 
We extracted the radius of curvature, $R$, by fitting a circle to  the polymer contour.   
Fig.~\ref{fig:R_of_fp}(a) shows the radius of curvature as a function of the 
propulsion strength rescaled by the filament bending stiffness, $f_p/\kappa$, 
for  two different values of $\alpha$, 0.75 and 0.5. 
As the propulsion force increases, the radius of curvature $R$ decreases
(the curvature $C=1/R$ of the polymer increases), until a saturation value is attained
that depends on the end-tailoring parameter $\alpha$ -- the smaller $\alpha$ and hence the larger
the end effect, the smaller the final radius (the higher the curvature).
For intermediate driving, there is a scaling regime where 
$R\propto\left(\frac{f_p}{\kappa}\right)^{-1}$, which is expected
from slender rod bending theory \cite{landau-elasticity}. 
For very small driving, there seems to be another regime
where $R\propto\left(\frac{f_p}{\kappa}\right)^{-1/2}$,
but it will not be observable in the presence of, even small, noise
as shown in Fig.~\ref{fig:R_of_fp}(b) for $k_BT=0.01$.
 In the presence of noise,
the  fluctuations (here assumed to be thermal) induce an average bending
that can be estimated by equaling thermal energy and bending energy,
$k_BT\sim\frac{\kappa}{2}\frac{L}{R^2}$. From there the system
directly crosses over into to $R\propto\left(\frac{f_p}{\kappa}\right)^{-1}$ regime.

Fig.~\ref{fig:R_of_fp}(c) shows the dependence of the saturated curvature
for high propulsion ($R_\infty$) on the end tailoring parameter $\alpha$.
It can be rationalized by kinematics as follows.
Consider the fiber to propel along the $y$-axis and let $\theta$ be the angle
between a local tangent and the $x$-axis.
Considering one of the two ends, the balance of the following forces must hold:
the propulsion force $\alpha f_p \mathbf{n}(\theta)$, which is along
the local normal $\mathbf{n}(\theta)$, the friction force $-\xi {\bf v}_{cm}$
proportional to the c.o.m.~velocity of the fiber
in $y$-direction, and the tension
force along the local tangent $\mathbf{t}(\theta)$. 
 
Eliminating the tension from the $x$-, $y$-components of the force balance 
and assuming $\xi { \bf v}_{cm}\simeq f_p$ for high driving force
(for a detailed study on the relation between shape and velocity, see below), 
one gets
$0=-\alpha f_p+\xi { \bf v}_{cm} \cos(\theta)\simeq
-\alpha f_p+f_p\cos(\theta)$. 
For a circular arc 
the angle at the filament end is given by $\theta=L/2R$ and hence 
\begin{equation}
R_{\infty}=\frac{L}{2\arccos(\alpha)}\,.
\end{equation} 
For $\alpha=1$, the fiber is straight ($R_{\infty}=\infty$);
for $\alpha=0$ this implies $\pi R_{\infty}=L$, corresponding to a half-circle.
Fig.~\ref{fig:R_of_fp}(c) shows a fit to simulations,
confirming the behavior $R_\infty\propto1/\arccos(\alpha)$.

We also investigated the effect of polymer length. 
In Fig.~\ref{fig:R_of_fp}(d), we tried a full scaling collapse,
plotting $R/L$ over $B=\frac{f_p L^3}{\kappa}$ for three different lengths
($N=10, 25, 50$). While  the scaling is fulfilled for every length on its own, 
there are offsets in both the $f_p^{-1}$ scaling regime and the plateau values,
implying that the bending is not perfectly circular arc-like.
In fact, while the end-tailored propulsion force bends the entire contour of shorter filaments, 
for longer filaments it affects mostly the boundary regions and the fiber is less
curved in the center region.

Let us now discuss the interplay of shape and propulsion velocity.
Since the fibers curve and the local propulsion force is orthogonal to the local tangent,
bent fibers will move more slowly.
The end-tailoring parameter $\alpha$, the stiffness, and the length of the polymer then  
decide  upon the c.o.m.~velocity of the filament. 
As seen in Fig.~\ref{fig:alpha-lessthan2-msd-beta}(a), the center of mass velocity
$|{\bf v}_{cm}| $ is no longer $f_p/\xi$, but reduced by a factor $\beta(\alpha,\kappa,N)$. 
The value of  $\beta$ was extracted from the slope of a straight line fit
to $|v_{cm}|$ as a function of $f_p$, for a given $\kappa$ and $N$. 
The resulting $\beta$ for different values of $\alpha$ and $N$ are shown in 
Fig.~\ref{fig:alpha-lessthan2-msd-beta}(b) as a function of stiffness $\kappa$. 
For a given $\alpha$, due to the end-induced deformation propagating to the middle of filament, 
short and flexible filaments curve more and propel slower than longer filaments
that curve less and translate faster for the same value of propulsion strength.

This observed velocity reduction is mostly a kinematic, shape-induced effect,
as can be seen as follows.
The propulsion force is $f_p\mathbf{n}$, except for the
ends. 
Neglecting the details of the end effects, 
but including their main effect, i.e.~the curving of the fiber to an approximately circular arc, 
the total force in propulsion direction is given by
(using symmetry)
$F_{tot}=2f_p\int_0^{L/2} \cos\theta(s)\,ds$.
For a circular arc,  one has $\theta(s)=\theta_e-s/R$
with the end angle $\theta_e=\frac{L}{2R}$.
Evaluating $F_{tot}$, expanding in $L/(2R)$
and equaling to the total friction force 
yields for the c.o.m.~velocity
\begin{equation}\label{veq}
 {\mathbf{v}_{cm}}=\frac{ f_p}{\xi}  \left[ 1 - c\left(\frac{L}{R}\right)^2     \right].
\end{equation}
Here $c$ is a constant (in this crude approximation $c=\frac{1}{24}$) 
and the radius of curvature $R=R(\alpha,\kappa,f_p)$ is a complicated function
of end-tailoring, stiffness and propulsion as investigated in Fig.~\ref{fig:R_of_fp}.
Note that, if $R$ were not dependent on the propulsion force $f_p$, the term 
in brackets in Eq.~(\ref{veq})
would just be $\beta$.

If indeed  the velocity reduction is due to the curving via simple kinematics, 
plotting $1-\frac{\mathbf{v}_{cm}}{f_p/\xi}$ over $(L/R)^2$ should yield a constant.
Fig.~\ref{fig:alpha-lessthan2-msd-beta}(c) shows the result:
first, there is a small offset, due to the 
reduced propulsion (due to $\alpha<1$) at the ends.
In fact, in the shown case 2 out of 25 beads are propelled 
with a fraction $\alpha=0.5$ of $f_p$ only,
which reduces the overall propulsion velocity by $0.04$. 
For finite $(L/R)^2$ there is a quite weak dependence
(of the same order as the end effect).  Taking into account that the scaling,
containing data from several orders of magnitude of both $\kappa$ and $f_p$,
works well shows that the kinematics of circular arcs gives a rather good description. 
Note that one can also apply the reasoning suggested by Eq.~(\ref{veq}) locally: 
wherever there is a curvature along the polymer contour, 
this leads to a reduction of the local velocity. This simple point of view will be useful 
for more complex polymer morphologies, cf.~the next section.


\section{Faster end beads ($\alpha>1$) and complex dynamics}

Finally, we consider the case of $\alpha>1$, where the end beads move faster 
than the polymer backbone. The distribution of the active propulsion then leads to 
a curving of the filament in the direction opposite to propulsion, 
see Fig~\ref{fig:schematic-alpha-conformation}b 
and Suppl.~movie 2.
As can be seen there, for weak end-tailoring $\alpha>1$, the shape is
just a mirror image of the case $\alpha<1$. However, for stronger $\alpha>1$, 
the behaviour becomes much more complex than for the case just described,
with strong deviations from the circular arc shape and even an instability
to oscillatory, flapping-like dynamics.

\begin{figure}
\includegraphics[width=0.4\textwidth]{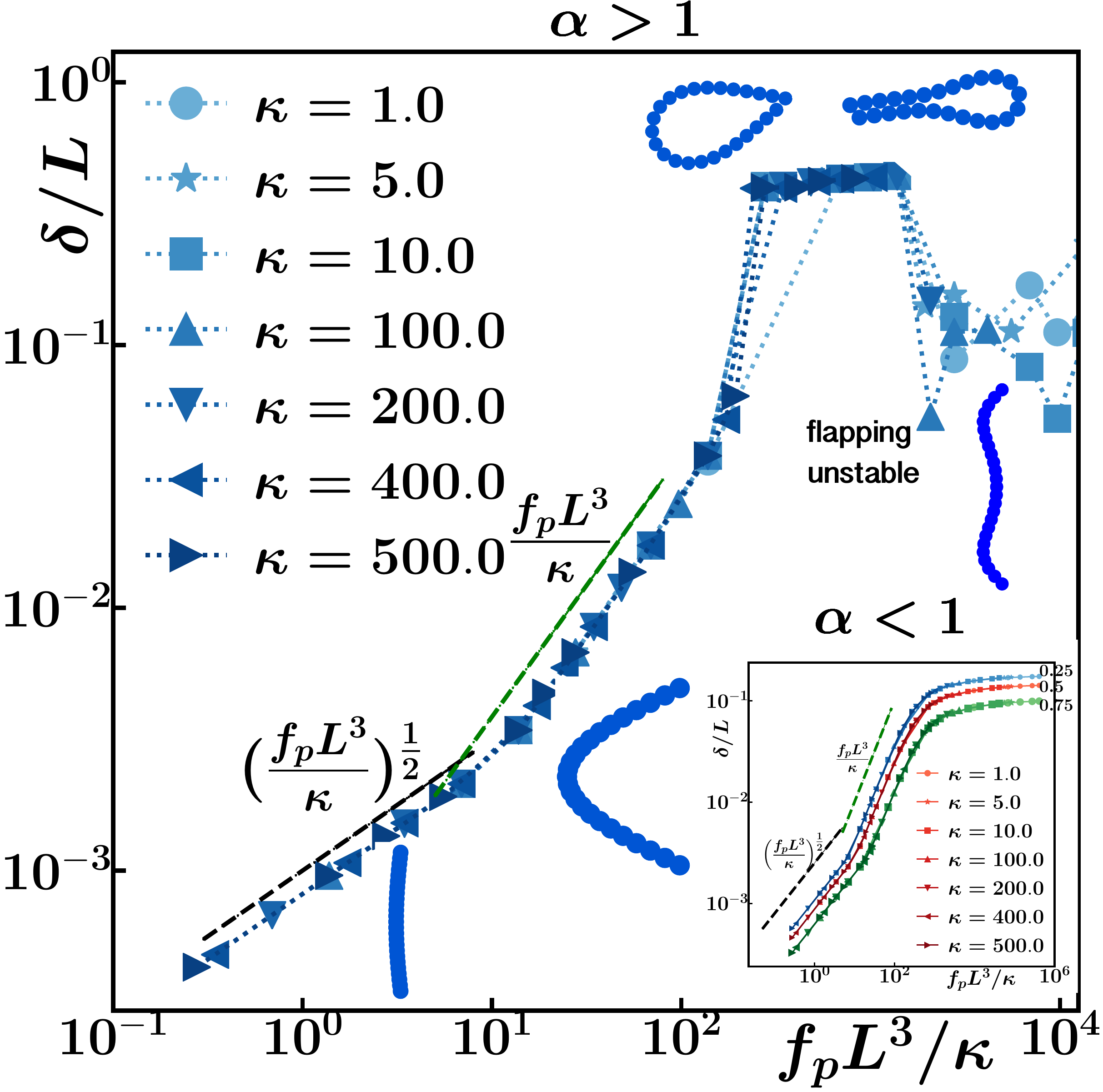}
\caption { Shape deformations for $\alpha>1$.
The scaled maximum amplitude of deformation $(\delta/L)$ is shown  
as a function of $B=f_pL^3/\kappa$ for $\alpha = 1.5$.  
Except for the dynamic states at high $B$, the data for different stiffnesses
collapse. Selected polymer conformations in the different regimes are presented. 
For comparison, the inset displays data for the case $\alpha<1$.
}
\label{fig:delta_beta_prime}
\end{figure}

To characterise these bent shapes, instead of the radius of curvature,
we hence opted for   the bending amplitude $\delta$, 
defined as the distance to the maximum curved point on the polymer contour 
(${\bf r}_{Cmax}$) from the center of mass of the end beads 
(${\bf r}_{cm}^ {1,N}$), 
\begin{equation}
\delta=|{\bf r}_{Cmax}^{} - {\bf r}_{cm}^ {1,N}|.
\end{equation}  
Fig~\ref{fig:delta_beta_prime} traces the scaled version $\delta/L$ as a function of 
$B=f_pL^3/\kappa$ together with corresponding
typical steady-state conformations. For comparison, the same plot for the case
$\alpha<1$ is given in the inset.

One can see from Fig~\ref{fig:delta_beta_prime} 
 that the scaling regimes, corresponding to those observed in the radius of
curvature in Fig.~\ref{fig:R_of_fp}, are again visible. However, in the case $\alpha>1$
there is not just saturation of curving (as for $\alpha<1$, cf.~the inset),
but complex shapes and even dynamic behavior can emerge:

At $B\gtrsim250$ for the given $\alpha=1.5$, a sudden jump in  $\delta$ occurs. 
This transition can be ascribed to the emergence of 
 \textit{2Dloop} or \textit{hairpin}  structures, 
 which form when the end beads of the polymer come into contact with each other,
 cf.~the snapshots in Fig~\ref{fig:delta_beta_prime} and Suppl.~movie 3 and 4, respectively.
 Once they fold on to each other, 
the self-avoiding interaction comes into play, but the larger propulsion
force of the ends keeps the structure together.
Typically, one end bead sits at the void between two beads 
of the opposite end, forming a bead-triplet configuration. 
Either a drop-shaped \textit{2Dloop} or a \textit{hairpin} conformation  is formed,
where in the latter several beads close to the ends form a quasi 1D 
triangular lattice-like arrangement, 
see the snapshots in Fig.~\ref{fig:delta_beta_prime}. 
While establishing the final shape, the polymer is almost stationary. 
The final polymer conformation, however, exhibits a stable rotational motion, 
since opposing beads cancel their propulsion, but only partially for reasons of shape asymmetry.
Hence the center of mass of both the $2D loop$ and $hairpin$ conformation 
then follow a circular path, the period of rotation being mostly determined by $f_p$.
%

\begin{figure}[]
\includegraphics[width=0.5\textwidth]{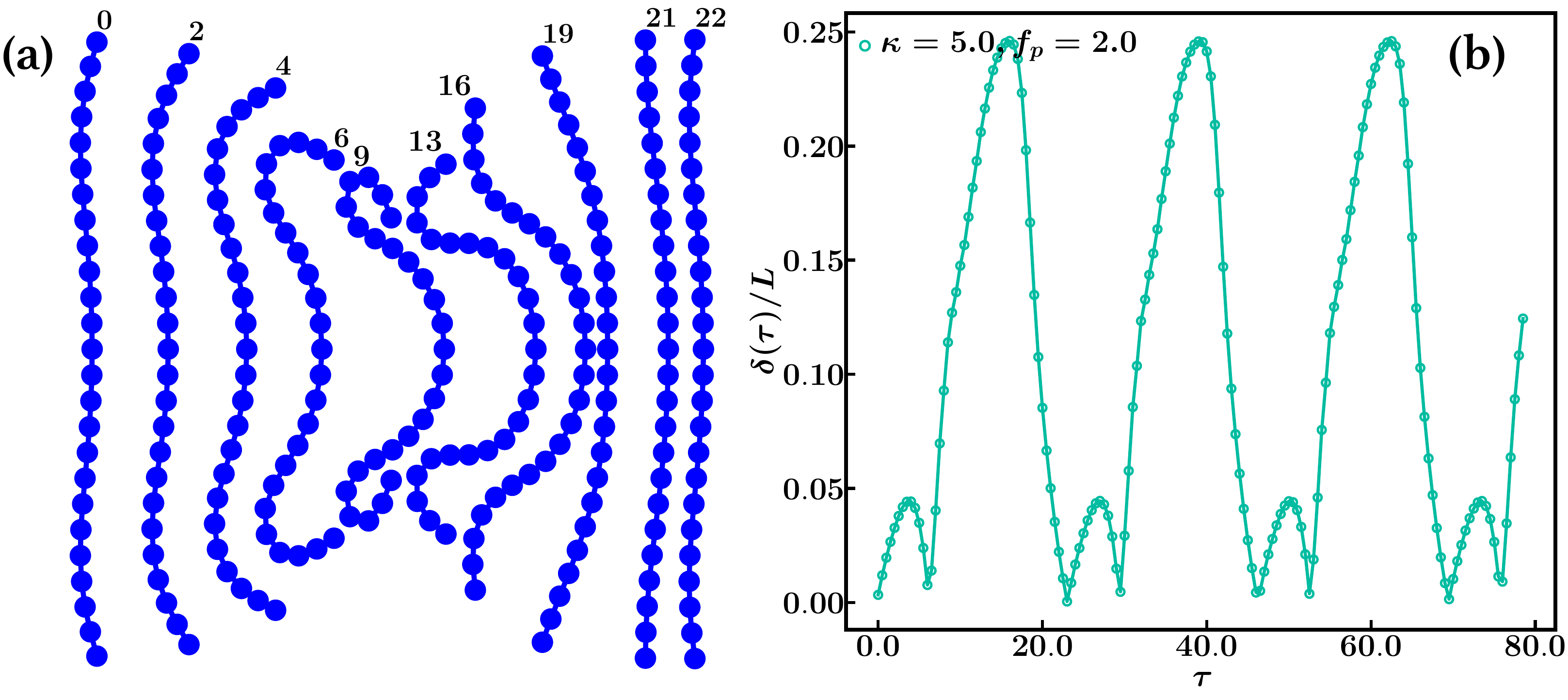}
\caption { Flapping motion for $\alpha>1$ and large $B$ values.
(a) A series of conformations is shown during one period 
(the resulting period is $23\tau$) 
of the flapping instability.
The fiber propells to the right.
Parameters:  $\alpha=1.5$ and  $B\simeq5530$ ($\kappa=5$, $f_p=2$, $L=25$). 
(b) Several oscillation periods 
are shown for  the scaled maximum amplitude of deformation, $\delta/L$, corresponding
to the shapes shown in (a).} 
\label{fig:flapping}
\end{figure}

At even higher drive, $B\gtrsim1600$ for the given $\alpha=1.5$,
an instability occurs, visible in Fig.~\ref{fig:delta_beta_prime} by  a drop of $\delta$.
Here an initially straight polymer forms a {\it W-shaped} conformation
that is highly dynamic, with the ends flapping in an oscillatory fashion.
A time-lapse of the conformational changes of such a cycle 
is shown in Fig.~\ref{fig:flapping}(a).
As can be seen the ends, being faster, ``roll up''.
In doing so, however, they contribute less to propulsion,
hence the initially rather straight middle catches up and starts to bulge out.
This curving, however, slows the middle down,
cf.~Eq.~(\ref{veq}) and the concomitant elastic deformation leads to an 
unfolding of the ends 
and finally the polymer attains its initial  {\it W-shape} again,
see also Suppl.~movie 5.

Note that in Fig.~\ref{fig:delta_beta_prime}, we always started from a straight initial conformation
to which the propulsion force was added. We also investigated whether transitions between the
different types of conformations are possible when changing the parameters 
$\alpha$ and/or $f_p$ for already established conformations.
Transitions between {\it 2Dloop} and {\it hairpin} are possible,
but if a fiber had already a closed {\it 2Dloop} or {\it hairpin} shape
it was not possible to directly induce the flapping {\it W-shape} by small continuous
parameter changes, i.e.~this state seems to be accessible only from initially roughly straight conformations.

From the bent shapes observed  in Fig.~\ref{fig:delta_beta_prime}
one would again expect a reduction in velocity
compared to straight fibers. For small deviations from $\alpha=1$, 
the system is mirror symmetric, cf. the snapshots in Fig.~\ref{fig:schematic-alpha-conformation},
and the velocity reduction for $\alpha>1$ is as for $\alpha<1$,
i.e.~roughly given by Eq.~(\ref{veq}),
albeit with an offset of opposite sign due to the stronger propulsion
of the end beads. For the more complex shapes,
\textit{2Dloop}, \textit{hairpin} and {\it W-shape},
the velocity is even further reduced, since more beads are pointing in directions
not contributing to propulsion or due to the oscillatory dynamics, respectively.

\begin{figure}
\includegraphics[width=0.4\textwidth]{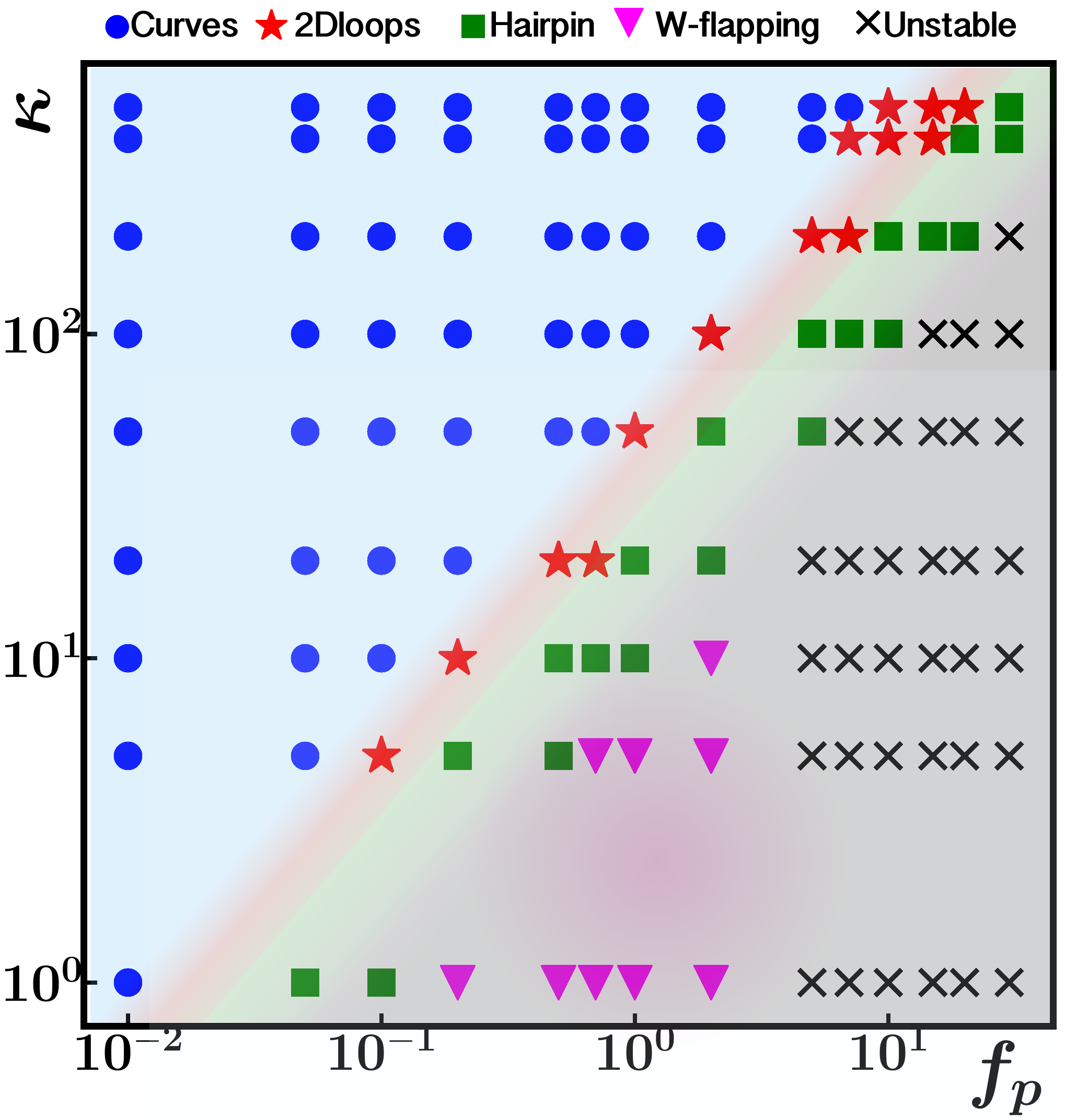}
\caption { State diagram in the $\kappa{-}f_p$ plane for $\alpha=1.5$, $N=25$.  
The different conformations were identified by their values of $\delta/L$, 
end-to-end distance $R_E$ and by visually checking. 
Crossovers between states  are indicated by shading  and serve as a 
guide to the eye.}
\label{fig:statediagram}
\end{figure}

Finally Fig.~\ref{fig:statediagram} displays a state diagram of the different
conformations attained by an initially straight polymer with $\alpha=1.5$ and $N=25$,
in the plane stiffness $\kappa$ vs.~propulsion force $f_p$. 
 For different lengths, such $\kappa{-}f_p$ diagrams look qualitatively the same, 
 but  overall there is no good data collapse using scaling,
 for the same reason as in the discussion of Fig.~\ref{fig:R_of_fp}. 
Also, for very short filaments ($N=5$), having too few degrees of freedom,
the complex conformations like the flapping are completely absent.
Nevertheless, as can be seen in Fig~\ref{fig:statediagram}, as the stiffness increases 
the simple curved conformations dominate
and the $2Dloop/hairpin$ conformations are shifted to higher values of $f_p$. 
The flapping mode occurs at low values of $\kappa$ and moderate value of $f_p$. 
Very large driving forces $f_p$ promote asymmetric polymer conformations
that are erratic/chaotic in time, denoted in the diagram as ``unstable''
and indicating that the system cannot adopt anymore simple, stable or time-periodic, states.

\section{Summary and outlook}

In this work, we used Langevin dynamics simulations to study the conformational properties 
of an active semi-flexible polymer that experiences local propulsion forces in the direction perpendicular 
to its contour. In addition the polymer is end-tailored, i.e.~the end beads have
slightly different propulsion properties.
We considered self-avoiding interactions but for simplicity assumed the dry limit, 
i.e.~solvent-mediated hydrodynamic interactions are ignored and friction is isotropic. 

We have shown that the conformational properties can be tuned by 
the relative propulsion strength between the end beads and the polymer backbone
and that a rich state diagram emerges from the interaction of 
propulsion and flexibility. 
If the ends beads move slower than the backbone (end-tailoring parameter $\alpha<1$), 
the polymer deforms to a bent conformation, 
and the curvature of the polymer is along the propulsion direction. 
The radius of curvature/the deformation amplitude shows scaling regimes 
due to the interplay of bending and active forces and a saturation regime 
at high propulsion where further increasing the propulsion force only stretches the polymer. 
Dynamic shapes as for $\alpha<1$ share similarities with bent conformations observed for moving 
filaments under a uniform field \cite{Ignacio_filamentdeformation_PRL_05,filamentgravity_exp_marchettiPRF_19} 
and with shapes of  rolling fibers driven by an energy flow through their cross-section
leading to a differential contraction \cite{Falko_fiberiod-2020}.   
  
When the end beads move faster than the backbone ($\alpha>1$), 
we have shown that different dynamic conformations emerge
as a function of stiffness and propulsion strength. 
For small propulsion a bent conformation with a curvature opposite to the propulsion direction is obtained.
For intermediate propulsion strength, the polymer can close into a rotating {\it 2Dloop} or {\it hairpin}
conformation.
Finally, for even larger driving the interplay between activity and flexibility 
can  trigger a dynamic  state of the polymer, which propells sideways 
while flapping around a {\it W-shape} conformation. 
Bent shapes as for $\alpha>1$
have been observed for rolling fibers driven by an energy flow through their cross-section
leading to differential expansion \cite{Falko_fiberiod-2020}.
Dynamic {\it W-shapes} were previously observed 
for sedimenting filaments in a viscous fluid \cite{Ignacio_filamentdeformation_PRL_05}.
Nevertheless, here the situation is different, because the propulsion
is in the transverse direction locally along the fiber shape, unlike for sedimentation.

We studied here only a single active polymer. 
In the last few years various collective effects have been extensively 
studied for ensembles of tangentially propelled objects
\cite{peruanibaer,vschaller-10,marchetti-13,prathyusha_PRE-18,Gompper-collective-polymer-18}.
In the context of their pattern formation and collective behaviour, however,
transversely propelled objects should belong to a different symmetry class 
as the tangential ones. As more and more such transverse propellers are being  designed
\cite{pietro-sagues-small10,vutukuri-huck-collective-transverse-rods-SM16,vrao-clasen-jphys-D-19,3d_printinghelix} it will be interesting to investigate their possibly different collective effects
in the future. Note that again flexibility will be important: namely,
 transversely propelled rigid colloidal janus rods were shown to rather block each other leading
to clustering and arrest of motion \cite{vutukuri-huck-collective-transverse-rods-SM16}.
In contrast, self-rolling flexible fibers were shown to be able to switch direction
by defoming under collisions \cite{Falko_fiberiod-2020}, which allows for more 
interesting collective behavior.
  
On the application side, sideways propellers have been already 
suggested to catch particles \cite{janussweeper-21}. Especially,
catalase-powered enzymatic Janus micromotors were already demonstrated 
to be able to capture bacteria \cite{motion-capture-bacterial-Linanoscale19}
and tumor cells \cite{fluroscene-tumour-Li-Chem.Eng.20}.
As we demonstrated here, including flexibility and adding ``end-tailoring''
substantially increases the possible conformations 
achievable for sideways propelled objects 
which in addition are dynamic, i.e. switchable via the fuel, 
hence opening up
new routes for the design of self-driven microrobotic engines.



\begin{thebibliography}{42}%
\makeatletter
\providecommand \@ifxundefined [1]{%
 \@ifx{#1\undefined}
}%
\providecommand \@ifnum [1]{%
 \ifnum #1\expandafter \@firstoftwo
 \else \expandafter \@secondoftwo
 \fi
}%
\providecommand \@ifx [1]{%
 \ifx #1\expandafter \@firstoftwo
 \else \expandafter \@secondoftwo
 \fi
}%
\providecommand \natexlab [1]{#1}%
\providecommand \enquote  [1]{``#1''}%
\providecommand \bibnamefont  [1]{#1}%
\providecommand \bibfnamefont [1]{#1}%
\providecommand \citenamefont [1]{#1}%
\providecommand \href@noop [0]{\@secondoftwo}%
\providecommand \href [0]{\begingroup \@sanitize@url \@href}%
\providecommand \@href[1]{\@@startlink{#1}\@@href}%
\providecommand \@@href[1]{\endgroup#1\@@endlink}%
\providecommand \@sanitize@url [0]{\catcode `\\12\catcode `\$12\catcode
  `\&12\catcode `\#12\catcode `\^12\catcode `\_12\catcode `\%12\relax}%
\providecommand \@@startlink[1]{}%
\providecommand \@@endlink[0]{}%
\providecommand \url  [0]{\begingroup\@sanitize@url \@url }%
\providecommand \@url [1]{\endgroup\@href {#1}{\urlprefix }}%
\providecommand \urlprefix  [0]{URL }%
\providecommand \Eprint [0]{\href }%
\providecommand \doibase [0]{http://dx.doi.org/}%
\providecommand \selectlanguage [0]{\@gobble}%
\providecommand \bibinfo  [0]{\@secondoftwo}%
\providecommand \bibfield  [0]{\@secondoftwo}%
\providecommand \translation [1]{[#1]}%
\providecommand \BibitemOpen [0]{}%
\providecommand \bibitemStop [0]{}%
\providecommand \bibitemNoStop [0]{.\EOS\space}%
\providecommand \EOS [0]{\spacefactor3000\relax}%
\providecommand \BibitemShut  [1]{\csname bibitem#1\endcsname}%
\let\auto@bib@innerbib\@empty
\bibitem [{\citenamefont {Gompper}\ \emph {et~al.}(2020)\citenamefont
  {Gompper}, \citenamefont {Winkler}, \citenamefont {Speck}, \citenamefont
  {Solon}, \citenamefont {Nardini}, \citenamefont {Peruani}, \citenamefont
  {Löwen}, \citenamefont {Golestanian}, \citenamefont {Kaupp}, \citenamefont
  {Alvarez}, \citenamefont {Ki{\o}rboe}, \citenamefont {Lauga}, \citenamefont
  {Poon}, \citenamefont {DeSimone}, \citenamefont {Mui{\~{n}}os-Landin},
  \citenamefont {Fischer}, \citenamefont {Söker}, \citenamefont {Cichos},
  \citenamefont {Kapral}, \citenamefont {Gaspard}, \citenamefont {Ripoll},
  \citenamefont {Sagues}, \citenamefont {Doostmohammadi}, \citenamefont
  {Yeomans}, \citenamefont {Aranson}, \citenamefont {Bechinger}, \citenamefont
  {Stark}, \citenamefont {Hemelrijk}, \citenamefont {Nedelec}, \citenamefont
  {Sarkar}, \citenamefont {Aryaksama}, \citenamefont {Lacroix}, \citenamefont
  {Duclos}, \citenamefont {Yashunsky}, \citenamefont {Silberzan}, \citenamefont
  {Arroyo},\ and\ \citenamefont {Kale}}]{roadmap}%
  \BibitemOpen
  \bibfield  {author} {\bibinfo {author} {\bibfnamefont {G.}~\bibnamefont
  {Gompper}}, \bibinfo {author} {\bibfnamefont {R.~G.}\ \bibnamefont
  {Winkler}}, \bibinfo {author} {\bibfnamefont {T.}~\bibnamefont {Speck}},
  \bibinfo {author} {\bibfnamefont {A.}~\bibnamefont {Solon}}, \bibinfo
  {author} {\bibfnamefont {C.}~\bibnamefont {Nardini}}, \bibinfo {author}
  {\bibfnamefont {F.}~\bibnamefont {Peruani}}, \bibinfo {author} {\bibfnamefont
  {H.}~\bibnamefont {Löwen}}, \bibinfo {author} {\bibfnamefont
  {R.}~\bibnamefont {Golestanian}}, \bibinfo {author} {\bibfnamefont {U.~B.}\
  \bibnamefont {Kaupp}}, \bibinfo {author} {\bibfnamefont {L.}~\bibnamefont
  {Alvarez}}, \bibinfo {author} {\bibfnamefont {T.}~\bibnamefont {Ki{\o}rboe}},
  \bibinfo {author} {\bibfnamefont {E.}~\bibnamefont {Lauga}}, \bibinfo
  {author} {\bibfnamefont {W.~C.~K.}\ \bibnamefont {Poon}}, \bibinfo {author}
  {\bibfnamefont {A.}~\bibnamefont {DeSimone}}, \bibinfo {author}
  {\bibfnamefont {S.}~\bibnamefont {Mui{\~{n}}os-Landin}}, \bibinfo {author}
  {\bibfnamefont {A.}~\bibnamefont {Fischer}}, \bibinfo {author} {\bibfnamefont
  {N.~A.}\ \bibnamefont {Söker}}, \bibinfo {author} {\bibfnamefont
  {F.}~\bibnamefont {Cichos}}, \bibinfo {author} {\bibfnamefont
  {R.}~\bibnamefont {Kapral}}, \bibinfo {author} {\bibfnamefont
  {P.}~\bibnamefont {Gaspard}}, \bibinfo {author} {\bibfnamefont
  {M.}~\bibnamefont {Ripoll}}, \bibinfo {author} {\bibfnamefont
  {F.}~\bibnamefont {Sagues}}, \bibinfo {author} {\bibfnamefont
  {A.}~\bibnamefont {Doostmohammadi}}, \bibinfo {author} {\bibfnamefont
  {J.~M.}\ \bibnamefont {Yeomans}}, \bibinfo {author} {\bibfnamefont {I.~S.}\
  \bibnamefont {Aranson}}, \bibinfo {author} {\bibfnamefont {C.}~\bibnamefont
  {Bechinger}}, \bibinfo {author} {\bibfnamefont {H.}~\bibnamefont {Stark}},
  \bibinfo {author} {\bibfnamefont {C.~K.}\ \bibnamefont {Hemelrijk}}, \bibinfo
  {author} {\bibfnamefont {F.~J.}\ \bibnamefont {Nedelec}}, \bibinfo {author}
  {\bibfnamefont {T.}~\bibnamefont {Sarkar}}, \bibinfo {author} {\bibfnamefont
  {T.}~\bibnamefont {Aryaksama}}, \bibinfo {author} {\bibfnamefont
  {M.}~\bibnamefont {Lacroix}}, \bibinfo {author} {\bibfnamefont
  {G.}~\bibnamefont {Duclos}}, \bibinfo {author} {\bibfnamefont
  {V.}~\bibnamefont {Yashunsky}}, \bibinfo {author} {\bibfnamefont
  {P.}~\bibnamefont {Silberzan}}, \bibinfo {author} {\bibfnamefont
  {M.}~\bibnamefont {Arroyo}}, \ and\ \bibinfo {author} {\bibfnamefont
  {S.}~\bibnamefont {Kale}},\ }\href@noop {} {\bibfield  {journal} {\bibinfo
  {journal} {J. Phys.: Condens. Matter}\ }\textbf {\bibinfo {volume} {32}},\
  \bibinfo {pages} {193001} (\bibinfo {year} {2020})}\BibitemShut {NoStop}%
\bibitem [{\citenamefont {Berg}\ and\ \citenamefont {Anderson}(1973)}]{Berg73}%
  \BibitemOpen
  \bibfield  {author} {\bibinfo {author} {\bibfnamefont {H.~C.}\ \bibnamefont
  {Berg}}\ and\ \bibinfo {author} {\bibfnamefont {R.~A.}\ \bibnamefont
  {Anderson}},\ }\href@noop {} {\bibfield  {journal} {\bibinfo  {journal}
  {Nature}\ }\textbf {\bibinfo {volume} {245}},\ \bibinfo {pages} {380}
  (\bibinfo {year} {1973})}\BibitemShut {NoStop}%
\bibitem [{\citenamefont {Lauga}\ and\ \citenamefont
  {Powers}(2009)}]{Lauga2009}%
  \BibitemOpen
  \bibfield  {author} {\bibinfo {author} {\bibfnamefont {E.}~\bibnamefont
  {Lauga}}\ and\ \bibinfo {author} {\bibfnamefont {T.~R.}\ \bibnamefont
  {Powers}},\ }\href@noop {} {\bibfield  {journal} {\bibinfo  {journal} {Rep.
  Prog. Phys.}\ }\textbf {\bibinfo {volume} {72}},\ \bibinfo {pages} {096601}
  (\bibinfo {year} {2009})}\BibitemShut {NoStop}%
\bibitem [{\citenamefont {Mauriello}\ \emph {et~al.}(2010)\citenamefont
  {Mauriello}, \citenamefont {Mignot}, \citenamefont {Yang},\ and\
  \citenamefont {Zusman}}]{myxoglide}%
  \BibitemOpen
  \bibfield  {author} {\bibinfo {author} {\bibfnamefont {E.~M.~F.}\
  \bibnamefont {Mauriello}}, \bibinfo {author} {\bibfnamefont {T.}~\bibnamefont
  {Mignot}}, \bibinfo {author} {\bibfnamefont {Z.}~\bibnamefont {Yang}}, \ and\
  \bibinfo {author} {\bibfnamefont {D.~R.}\ \bibnamefont {Zusman}},\
  }\href@noop {} {\bibfield  {journal} {\bibinfo  {journal} {Microbiol. Mol.
  Biol. Rev.}\ }\textbf {\bibinfo {volume} {74}},\ \bibinfo {pages} {229}
  (\bibinfo {year} {2010})}\BibitemShut {NoStop}%
\bibitem [{\citenamefont {Peruani}\ \emph {et~al.}(2012)\citenamefont
  {Peruani}, \citenamefont {Starru{\ss}}, \citenamefont {Jakovljevic},
  \citenamefont {{Sogaard-Andersen}}, \citenamefont {Deutsch},\ and\
  \citenamefont {B{\"a}r}}]{peruanibaer}%
  \BibitemOpen
  \bibfield  {author} {\bibinfo {author} {\bibfnamefont {F.}~\bibnamefont
  {Peruani}}, \bibinfo {author} {\bibfnamefont {J.}~\bibnamefont
  {Starru{\ss}}}, \bibinfo {author} {\bibfnamefont {V.}~\bibnamefont
  {Jakovljevic}}, \bibinfo {author} {\bibfnamefont {L.}~\bibnamefont
  {{Sogaard-Andersen}}}, \bibinfo {author} {\bibfnamefont {A.}~\bibnamefont
  {Deutsch}}, \ and\ \bibinfo {author} {\bibfnamefont {M.}~\bibnamefont
  {B{\"a}r}},\ }\href@noop {} {\bibfield  {journal} {\bibinfo  {journal} {Phys.
  Rev. Lett.}\ }\textbf {\bibinfo {volume} {108}},\ \bibinfo {pages} {098102}
  (\bibinfo {year} {2012})}\BibitemShut {NoStop}%
\bibitem [{\citenamefont {Ray}\ \emph {et~al.}(1993)\citenamefont {Ray},
  \citenamefont {Meyhöfer}, \citenamefont {Milligan},\ and\ \citenamefont
  {Howard}}]{howardMTglide}%
  \BibitemOpen
  \bibfield  {author} {\bibinfo {author} {\bibfnamefont {S.}~\bibnamefont
  {Ray}}, \bibinfo {author} {\bibfnamefont {E.}~\bibnamefont {Meyhöfer}},
  \bibinfo {author} {\bibfnamefont {R.~A.}\ \bibnamefont {Milligan}}, \ and\
  \bibinfo {author} {\bibfnamefont {J.}~\bibnamefont {Howard}},\ }\href@noop {}
  {\bibfield  {journal} {\bibinfo  {journal} {J. Cell Biol.}\ }\textbf
  {\bibinfo {volume} {121}},\ \bibinfo {pages} {1083} (\bibinfo {year}
  {1993})}\BibitemShut {NoStop}%
\bibitem [{\citenamefont {Schaller}\ \emph {et~al.}(2010)\citenamefont
  {Schaller}, \citenamefont {Weber}, \citenamefont {Semmrich}, \citenamefont
  {Frey},\ and\ \citenamefont {Bausch}}]{vschaller-10}%
  \BibitemOpen
  \bibfield  {author} {\bibinfo {author} {\bibfnamefont {V.}~\bibnamefont
  {Schaller}}, \bibinfo {author} {\bibfnamefont {C.}~\bibnamefont {Weber}},
  \bibinfo {author} {\bibfnamefont {C.}~\bibnamefont {Semmrich}}, \bibinfo
  {author} {\bibfnamefont {E.}~\bibnamefont {Frey}}, \ and\ \bibinfo {author}
  {\bibfnamefont {A.~R.}\ \bibnamefont {Bausch}},\ }\href@noop {} {\bibfield
  {journal} {\bibinfo  {journal} {Nature}\ }\textbf {\bibinfo {volume} {467}},\
  \bibinfo {pages} {73} (\bibinfo {year} {2010})}\BibitemShut {NoStop}%
\bibitem [{\citenamefont {Paxton}\ \emph {et~al.}(2004)\citenamefont {Paxton},
  \citenamefont {Kistler}, \citenamefont {Olmeda}, \citenamefont {Sen},
  \citenamefont {Angelo}, \citenamefont {Cao}, \citenamefont {Mallouk},
  \citenamefont {Lammert},\ and\ \citenamefont {Crespi}}]{active_Pt_Aurods}%
  \BibitemOpen
  \bibfield  {author} {\bibinfo {author} {\bibfnamefont {W.~F.}\ \bibnamefont
  {Paxton}}, \bibinfo {author} {\bibfnamefont {K.~C.}\ \bibnamefont {Kistler}},
  \bibinfo {author} {\bibfnamefont {C.~C.}\ \bibnamefont {Olmeda}}, \bibinfo
  {author} {\bibfnamefont {A.}~\bibnamefont {Sen}}, \bibinfo {author}
  {\bibfnamefont {S.~K.~S.}\ \bibnamefont {Angelo}}, \bibinfo {author}
  {\bibfnamefont {Y.}~\bibnamefont {Cao}}, \bibinfo {author} {\bibfnamefont
  {T.~E.}\ \bibnamefont {Mallouk}}, \bibinfo {author} {\bibfnamefont {P.~E.}\
  \bibnamefont {Lammert}}, \ and\ \bibinfo {author} {\bibfnamefont {V.~H.}\
  \bibnamefont {Crespi}},\ }\href@noop {} {\bibfield  {journal} {\bibinfo
  {journal} {J. Am. Chem. Soc.}\ }\textbf {\bibinfo {volume} {126}},\ \bibinfo
  {pages} {13431} (\bibinfo {year} {2004})}\BibitemShut {NoStop}%
\bibitem [{\citenamefont {Ebbens}\ \emph {et~al.}(2012)\citenamefont {Ebbens},
  \citenamefont {Tu}, \citenamefont {Howse},\ and\ \citenamefont
  {Golestanian}}]{Ebbens-Ramin-PRE-12}%
  \BibitemOpen
  \bibfield  {author} {\bibinfo {author} {\bibfnamefont {S.}~\bibnamefont
  {Ebbens}}, \bibinfo {author} {\bibfnamefont {M.-H.}\ \bibnamefont {Tu}},
  \bibinfo {author} {\bibfnamefont {J.~R.}\ \bibnamefont {Howse}}, \ and\
  \bibinfo {author} {\bibfnamefont {R.}~\bibnamefont {Golestanian}},\
  }\href@noop {} {\bibfield  {journal} {\bibinfo  {journal} {Phys. Rev. E}\
  }\textbf {\bibinfo {volume} {85}},\ \bibinfo {pages} {020401(R)} (\bibinfo
  {year} {2012})}\BibitemShut {NoStop}%
\bibitem [{\citenamefont {Howse}\ \emph {et~al.}(2007)\citenamefont {Howse},
  \citenamefont {Jones}, \citenamefont {Ryan}, \citenamefont {Gough},
  \citenamefont {Vafabakhsh},\ and\ \citenamefont {Golestanian}}]{HowseRamin}%
  \BibitemOpen
  \bibfield  {author} {\bibinfo {author} {\bibfnamefont {J.~R.}\ \bibnamefont
  {Howse}}, \bibinfo {author} {\bibfnamefont {R.~A.~L.}\ \bibnamefont {Jones}},
  \bibinfo {author} {\bibfnamefont {A.~J.}\ \bibnamefont {Ryan}}, \bibinfo
  {author} {\bibfnamefont {T.}~\bibnamefont {Gough}}, \bibinfo {author}
  {\bibfnamefont {R.}~\bibnamefont {Vafabakhsh}}, \ and\ \bibinfo {author}
  {\bibfnamefont {R.}~\bibnamefont {Golestanian}},\ }\href@noop {} {\bibfield
  {journal} {\bibinfo  {journal} {Phys. Rev. Lett.}\ }\textbf {\bibinfo
  {volume} {99}},\ \bibinfo {pages} {048102} (\bibinfo {year}
  {2007})}\BibitemShut {NoStop}%
\bibitem [{\citenamefont {Soto}\ and\ \citenamefont
  {Golestanian}(2014)}]{Soto-Ramin-PRL-14}%
  \BibitemOpen
  \bibfield  {author} {\bibinfo {author} {\bibfnamefont {R.}~\bibnamefont
  {Soto}}\ and\ \bibinfo {author} {\bibfnamefont {R.}~\bibnamefont
  {Golestanian}},\ }\href@noop {} {\bibfield  {journal} {\bibinfo  {journal}
  {Phys. Rev. Lett.}\ }\textbf {\bibinfo {volume} {112}},\ \bibinfo {pages}
  {068301} (\bibinfo {year} {2014})}\BibitemShut {NoStop}%
\bibitem [{\citenamefont {Buttinoni}\ \emph {et~al.}(2012)\citenamefont
  {Buttinoni}, \citenamefont {Volpe}, \citenamefont {Kümmel}, \citenamefont
  {Volpe},\ and\ \citenamefont {Bechinger}}]{BechingerJPCM}%
  \BibitemOpen
  \bibfield  {author} {\bibinfo {author} {\bibfnamefont {I.}~\bibnamefont
  {Buttinoni}}, \bibinfo {author} {\bibfnamefont {G.}~\bibnamefont {Volpe}},
  \bibinfo {author} {\bibfnamefont {F.}~\bibnamefont {Kümmel}}, \bibinfo
  {author} {\bibfnamefont {G.}~\bibnamefont {Volpe}}, \ and\ \bibinfo {author}
  {\bibfnamefont {C.}~\bibnamefont {Bechinger}},\ }\href@noop {} {\bibfield
  {journal} {\bibinfo  {journal} {J. Phys. Cond. Matt.}\ }\textbf {\bibinfo
  {volume} {24}},\ \bibinfo {pages} {284129} (\bibinfo {year}
  {2012})}\BibitemShut {NoStop}%
\bibitem [{\citenamefont {Tierno}\ \emph {et~al.}(2010)\citenamefont {Tierno},
  \citenamefont {Albalat},\ and\ \citenamefont
  {Sagues}}]{pietro-sagues-small10}%
  \BibitemOpen
  \bibfield  {author} {\bibinfo {author} {\bibfnamefont {P.}~\bibnamefont
  {Tierno}}, \bibinfo {author} {\bibfnamefont {R.}~\bibnamefont {Albalat}}, \
  and\ \bibinfo {author} {\bibfnamefont {F.}~\bibnamefont {Sagues}},\
  }\href@noop {} {\bibfield  {journal} {\bibinfo  {journal} {Small}\ }\textbf
  {\bibinfo {volume} {16}},\ \bibinfo {pages} {1749} (\bibinfo {year}
  {2010})}\BibitemShut {NoStop}%
\bibitem [{\citenamefont {Vutukuri}\ \emph {et~al.}(2016)\citenamefont
  {Vutukuri}, \citenamefont {Preisler}, \citenamefont {Besseling},
  \citenamefont {{van Blaaderen}}, \citenamefont {Dijkstra},\ and\
  \citenamefont {Huck}}]{vutukuri-huck-collective-transverse-rods-SM16}%
  \BibitemOpen
  \bibfield  {author} {\bibinfo {author} {\bibfnamefont {H.~R.}\ \bibnamefont
  {Vutukuri}}, \bibinfo {author} {\bibfnamefont {Z.}~\bibnamefont {Preisler}},
  \bibinfo {author} {\bibfnamefont {T.~H.}\ \bibnamefont {Besseling}}, \bibinfo
  {author} {\bibfnamefont {A.}~\bibnamefont {{van Blaaderen}}}, \bibinfo
  {author} {\bibfnamefont {M.}~\bibnamefont {Dijkstra}}, \ and\ \bibinfo
  {author} {\bibfnamefont {W.~T.~S.}\ \bibnamefont {Huck}},\ }\href@noop {}
  {\bibfield  {journal} {\bibinfo  {journal} {Soft Matter}\ }\textbf {\bibinfo
  {volume} {12}},\ \bibinfo {pages} {9657} (\bibinfo {year}
  {2016})}\BibitemShut {NoStop}%
\bibitem [{\citenamefont {Rao}\ \emph {et~al.}(2019)\citenamefont {Rao},
  \citenamefont {Reddy}, \citenamefont {Fransaer},\ and\ \citenamefont
  {Clasen}}]{vrao-clasen-jphys-D-19}%
  \BibitemOpen
  \bibfield  {author} {\bibinfo {author} {\bibfnamefont {D.~V.}\ \bibnamefont
  {Rao}}, \bibinfo {author} {\bibfnamefont {N.}~\bibnamefont {Reddy}}, \bibinfo
  {author} {\bibfnamefont {J.}~\bibnamefont {Fransaer}}, \ and\ \bibinfo
  {author} {\bibfnamefont {C.}~\bibnamefont {Clasen}},\ }\href@noop {}
  {\bibfield  {journal} {\bibinfo  {journal} {J. Phys. D. Apply. Phys.}\
  }\textbf {\bibinfo {volume} {52}},\ \bibinfo {pages} {014002} (\bibinfo
  {year} {2019})}\BibitemShut {NoStop}%
\bibitem [{\citenamefont {Doherty}\ \emph {et~al.}(2020)\citenamefont
  {Doherty}, \citenamefont {Varkevisser}, \citenamefont {Teunisse},
  \citenamefont {Hoecht}, \citenamefont {Ketzetzi}, \citenamefont {Ouhajji},\
  and\ \citenamefont {Kraft}}]{3d_printinghelix}%
  \BibitemOpen
  \bibfield  {author} {\bibinfo {author} {\bibfnamefont {R.~P.}\ \bibnamefont
  {Doherty}}, \bibinfo {author} {\bibfnamefont {T.}~\bibnamefont
  {Varkevisser}}, \bibinfo {author} {\bibfnamefont {M.}~\bibnamefont
  {Teunisse}}, \bibinfo {author} {\bibfnamefont {J.}~\bibnamefont {Hoecht}},
  \bibinfo {author} {\bibfnamefont {S.}~\bibnamefont {Ketzetzi}}, \bibinfo
  {author} {\bibfnamefont {S.}~\bibnamefont {Ouhajji}}, \ and\ \bibinfo
  {author} {\bibfnamefont {D.~J.}\ \bibnamefont {Kraft}},\ }\href@noop {}
  {\bibfield  {journal} {\bibinfo  {journal} {Soft Matter}\ }\textbf {\bibinfo
  {volume} {16}},\ \bibinfo {pages} {10463} (\bibinfo {year}
  {2020})}\BibitemShut {NoStop}%
\bibitem [{\citenamefont {Vutukuri}\ \emph {et~al.}(2017)\citenamefont
  {Vutukuri}, \citenamefont {Bet}, \citenamefont {van Roij}, \citenamefont
  {Dijkstra},\ and\ \citenamefont {Huck}}]{vrao-huck-scirepD-17}%
  \BibitemOpen
  \bibfield  {author} {\bibinfo {author} {\bibfnamefont {H.~R.}\ \bibnamefont
  {Vutukuri}}, \bibinfo {author} {\bibfnamefont {B.}~\bibnamefont {Bet}},
  \bibinfo {author} {\bibfnamefont {R.}~\bibnamefont {van Roij}}, \bibinfo
  {author} {\bibfnamefont {M.}~\bibnamefont {Dijkstra}}, \ and\ \bibinfo
  {author} {\bibfnamefont {W.~T.~S.}\ \bibnamefont {Huck}},\ }\href@noop {}
  {\bibfield  {journal} {\bibinfo  {journal} {Sci. Rep}\ }\textbf {\bibinfo
  {volume} {7}},\ \bibinfo {pages} {16758} (\bibinfo {year}
  {2017})}\BibitemShut {NoStop}%
\bibitem [{\citenamefont {Baumann}\ \emph {et~al.}(2018)\citenamefont
  {Baumann}, \citenamefont {Sanchez-Ferrer}, \citenamefont {Jacomine},
  \citenamefont {Martinoty}, \citenamefont {Houerou}, \citenamefont {Ziebert},\
  and\ \citenamefont {Kulic}}]{falko-nature-material-18}%
  \BibitemOpen
  \bibfield  {author} {\bibinfo {author} {\bibfnamefont {A.}~\bibnamefont
  {Baumann}}, \bibinfo {author} {\bibfnamefont {A.}~\bibnamefont
  {Sanchez-Ferrer}}, \bibinfo {author} {\bibfnamefont {L.}~\bibnamefont
  {Jacomine}}, \bibinfo {author} {\bibfnamefont {P.}~\bibnamefont {Martinoty}},
  \bibinfo {author} {\bibfnamefont {V.~L.}\ \bibnamefont {Houerou}}, \bibinfo
  {author} {\bibfnamefont {F.}~\bibnamefont {Ziebert}}, \ and\ \bibinfo
  {author} {\bibfnamefont {I.~M.}\ \bibnamefont {Kulic}},\ }\href@noop {}
  {\bibfield  {journal} {\bibinfo  {journal} {Nat. mater.}\ }\textbf {\bibinfo
  {volume} {17}},\ \bibinfo {pages} {523} (\bibinfo {year} {2018})}\BibitemShut
  {NoStop}%
\bibitem [{\citenamefont {Bazir}\ \emph {et~al.}(2020)\citenamefont {Bazir},
  \citenamefont {Baumann}, \citenamefont {Ziebert},\ and\ \citenamefont
  {Kulic}}]{Falko_fiberiod-2020}%
  \BibitemOpen
  \bibfield  {author} {\bibinfo {author} {\bibfnamefont {A.}~\bibnamefont
  {Bazir}}, \bibinfo {author} {\bibfnamefont {A.}~\bibnamefont {Baumann}},
  \bibinfo {author} {\bibfnamefont {F.}~\bibnamefont {Ziebert}}, \ and\
  \bibinfo {author} {\bibfnamefont {I.~M.}\ \bibnamefont {Kulic}},\ }\href@noop
  {} {\bibfield  {journal} {\bibinfo  {journal} {Soft Matter}\ }\textbf
  {\bibinfo {volume} {16}},\ \bibinfo {pages} {5210} (\bibinfo {year}
  {2020})}\BibitemShut {NoStop}%
\bibitem [{\citenamefont {Jiang}\ \emph {et~al.}(2021)\citenamefont {Jiang},
  \citenamefont {Xiao}, \citenamefont {Cheng}, \citenamefont {Hou},\ and\
  \citenamefont {Zhao}}]{rolling-spring-actuator-ChemMat-21}%
  \BibitemOpen
  \bibfield  {author} {\bibinfo {author} {\bibfnamefont {Z.-C.}\ \bibnamefont
  {Jiang}}, \bibinfo {author} {\bibfnamefont {Y.-Y.}\ \bibnamefont {Xiao}},
  \bibinfo {author} {\bibfnamefont {R.-D.}\ \bibnamefont {Cheng}}, \bibinfo
  {author} {\bibfnamefont {J.-B.}\ \bibnamefont {Hou}}, \ and\ \bibinfo
  {author} {\bibfnamefont {Y.}~\bibnamefont {Zhao}},\ }\href@noop {} {\bibfield
   {journal} {\bibinfo  {journal} {Chem. Mater.}\ }\textbf {\bibinfo {volume}
  {33}},\ \bibinfo {pages} {6541} (\bibinfo {year} {2021})}\BibitemShut
  {NoStop}%
\bibitem [{\citenamefont {Yan}\ \emph {et~al.}(2016)\citenamefont {Yan},
  \citenamefont {Han}, \citenamefont {Zhang}, \citenamefont {Xu}, \citenamefont
  {Luijten},\ and\ \citenamefont {Granick}}]{granick-janus-chain-nat-mat16}%
  \BibitemOpen
  \bibfield  {author} {\bibinfo {author} {\bibfnamefont {J.}~\bibnamefont
  {Yan}}, \bibinfo {author} {\bibfnamefont {M.}~\bibnamefont {Han}}, \bibinfo
  {author} {\bibfnamefont {J.}~\bibnamefont {Zhang}}, \bibinfo {author}
  {\bibfnamefont {C.}~\bibnamefont {Xu}}, \bibinfo {author} {\bibfnamefont
  {E.}~\bibnamefont {Luijten}}, \ and\ \bibinfo {author} {\bibfnamefont
  {S.}~\bibnamefont {Granick}},\ }\href@noop {} {\bibfield  {journal} {\bibinfo
   {journal} {Nat. Mater.}\ }\textbf {\bibinfo {volume} {15}},\ \bibinfo
  {pages} {1095} (\bibinfo {year} {2016})}\BibitemShut {NoStop}%
\bibitem [{\citenamefont {Jayaraman}\ \emph {et~al.}(2012)\citenamefont
  {Jayaraman}, \citenamefont {Ramachandran}, \citenamefont {Ghose},
  \citenamefont {Laskar}, \citenamefont {Bhamla}, \citenamefont {Kumar},\ and\
  \citenamefont {Adhikari}}]{gayathri-12}%
  \BibitemOpen
  \bibfield  {author} {\bibinfo {author} {\bibfnamefont {G.}~\bibnamefont
  {Jayaraman}}, \bibinfo {author} {\bibfnamefont {S.}~\bibnamefont
  {Ramachandran}}, \bibinfo {author} {\bibfnamefont {S.}~\bibnamefont {Ghose}},
  \bibinfo {author} {\bibfnamefont {A.}~\bibnamefont {Laskar}}, \bibinfo
  {author} {\bibfnamefont {M.~S.}\ \bibnamefont {Bhamla}}, \bibinfo {author}
  {\bibfnamefont {P.~B.~S.}\ \bibnamefont {Kumar}}, \ and\ \bibinfo {author}
  {\bibfnamefont {R.}~\bibnamefont {Adhikari}},\ }\href@noop {} {\bibfield
  {journal} {\bibinfo  {journal} {Phys. Rev. Lett.}\ }\textbf {\bibinfo
  {volume} {109}},\ \bibinfo {pages} {158302} (\bibinfo {year}
  {2012})}\BibitemShut {NoStop}%
\bibitem [{\citenamefont {Isele-Holder}\ \emph {et~al.}(2015)\citenamefont
  {Isele-Holder}, \citenamefont {Elgeti},\ and\ \citenamefont
  {Gompper}}]{riseleholder-15}%
  \BibitemOpen
  \bibfield  {author} {\bibinfo {author} {\bibfnamefont {R.~E.}\ \bibnamefont
  {Isele-Holder}}, \bibinfo {author} {\bibfnamefont {J.}~\bibnamefont
  {Elgeti}}, \ and\ \bibinfo {author} {\bibfnamefont {G.}~\bibnamefont
  {Gompper}},\ }\href@noop {} {\bibfield  {journal} {\bibinfo  {journal} {Soft
  Matter}\ }\textbf {\bibinfo {volume} {11}},\ \bibinfo {pages} {7181}
  (\bibinfo {year} {2015})}\BibitemShut {NoStop}%
\bibitem [{\citenamefont {Eisenstecken}\ \emph {et~al.}(2016)\citenamefont
  {Eisenstecken}, \citenamefont {Gompper},\ and\ \citenamefont
  {Winkler}}]{Winkler-conformational-property-16}%
  \BibitemOpen
  \bibfield  {author} {\bibinfo {author} {\bibfnamefont {T.}~\bibnamefont
  {Eisenstecken}}, \bibinfo {author} {\bibfnamefont {G.}~\bibnamefont
  {Gompper}}, \ and\ \bibinfo {author} {\bibfnamefont {R.~G.}\ \bibnamefont
  {Winkler}},\ }\href@noop {} {\bibfield  {journal} {\bibinfo  {journal}
  {Polymers}\ }\textbf {\bibinfo {volume} {8}},\ \bibinfo {pages} {304}
  (\bibinfo {year} {2016})}\BibitemShut {NoStop}%
\bibitem [{\citenamefont {Jiang}\ and\ \citenamefont {Hou}(2014)}]{hiang-14a}%
  \BibitemOpen
  \bibfield  {author} {\bibinfo {author} {\bibfnamefont {H.}~\bibnamefont
  {Jiang}}\ and\ \bibinfo {author} {\bibfnamefont {Z.}~\bibnamefont {Hou}},\
  }\href@noop {} {\bibfield  {journal} {\bibinfo  {journal} {Soft Matter}\
  }\textbf {\bibinfo {volume} {10}},\ \bibinfo {pages} {1012} (\bibinfo {year}
  {2014})}\BibitemShut {NoStop}%
\bibitem [{\citenamefont {Lagomarsino}\ \emph {et~al.}(2005)\citenamefont
  {Lagomarsino}, \citenamefont {Pagonabarraga},\ and\ \citenamefont
  {Lowe}}]{Ignacio_filamentdeformation_PRL_05}%
  \BibitemOpen
  \bibfield  {author} {\bibinfo {author} {\bibfnamefont {M.~C.}\ \bibnamefont
  {Lagomarsino}}, \bibinfo {author} {\bibfnamefont {I.}~\bibnamefont
  {Pagonabarraga}}, \ and\ \bibinfo {author} {\bibfnamefont {C.~P.}\
  \bibnamefont {Lowe}},\ }\href@noop {} {\bibfield  {journal} {\bibinfo
  {journal} {Phys. Rev. Lett.}\ }\textbf {\bibinfo {volume} {94}},\ \bibinfo
  {pages} {148104} (\bibinfo {year} {2005})}\BibitemShut {NoStop}%
\bibitem [{\citenamefont {Xu}\ and\ \citenamefont
  {Nadim}(1994)}]{Ali_nadim_deformatin_orientation_phys_fluids_94}%
  \BibitemOpen
  \bibfield  {author} {\bibinfo {author} {\bibfnamefont {X.}~\bibnamefont
  {Xu}}\ and\ \bibinfo {author} {\bibfnamefont {A.}~\bibnamefont {Nadim}},\
  }\href@noop {} {\bibfield  {journal} {\bibinfo  {journal} {Phys. Fluids}\
  }\textbf {\bibinfo {volume} {9}},\ \bibinfo {pages} {2890} (\bibinfo {year}
  {1994})}\BibitemShut {NoStop}%
\bibitem [{\citenamefont {Schlagberger}\ and\ \citenamefont
  {Netz}(2005)}]{Netz_EPL_05}%
  \BibitemOpen
  \bibfield  {author} {\bibinfo {author} {\bibfnamefont {X.}~\bibnamefont
  {Schlagberger}}\ and\ \bibinfo {author} {\bibfnamefont {R.~R.}\ \bibnamefont
  {Netz}},\ }\href@noop {} {\bibfield  {journal} {\bibinfo  {journal} {Euro.
  Phys. Lett.}\ }\textbf {\bibinfo {volume} {70}},\ \bibinfo {pages} {129}
  (\bibinfo {year} {2005})}\BibitemShut {NoStop}%
\bibitem [{\citenamefont {Marchetti}\ \emph {et~al.}(2019)\citenamefont
  {Marchetti}, \citenamefont {Raspa}, \citenamefont {Lindner}, \citenamefont
  {du~Roure}, \citenamefont {Bergougnoux}, \citenamefont {Guazzelli},\ and\
  \citenamefont {Duprat}}]{filamentgravity_exp_marchettiPRF_19}%
  \BibitemOpen
  \bibfield  {author} {\bibinfo {author} {\bibfnamefont {B.}~\bibnamefont
  {Marchetti}}, \bibinfo {author} {\bibfnamefont {V.}~\bibnamefont {Raspa}},
  \bibinfo {author} {\bibfnamefont {A.}~\bibnamefont {Lindner}}, \bibinfo
  {author} {\bibfnamefont {O.}~\bibnamefont {du~Roure}}, \bibinfo {author}
  {\bibfnamefont {L.}~\bibnamefont {Bergougnoux}}, \bibinfo {author}
  {\bibfnamefont {{\' E}.}~\bibnamefont {Guazzelli}}, \ and\ \bibinfo {author}
  {\bibfnamefont {C.}~\bibnamefont {Duprat}},\ }\href@noop {} {\bibfield
  {journal} {\bibinfo  {journal} {Phys. Rev. Fluids}\ }\textbf {\bibinfo
  {volume} {3}},\ \bibinfo {pages} {104102} (\bibinfo {year}
  {2019})}\BibitemShut {NoStop}%
\bibitem [{\citenamefont {Weeks}\ \emph {et~al.}(1971)\citenamefont {Weeks},
  \citenamefont {Chandler},\ and\ \citenamefont
  {Andersen}}]{weeks-chandler-71}%
  \BibitemOpen
  \bibfield  {author} {\bibinfo {author} {\bibfnamefont {J.~D.}\ \bibnamefont
  {Weeks}}, \bibinfo {author} {\bibfnamefont {D.}~\bibnamefont {Chandler}}, \
  and\ \bibinfo {author} {\bibfnamefont {H.~C.}\ \bibnamefont {Andersen}},\
  }\href@noop {} {\bibfield  {journal} {\bibinfo  {journal} {J. Chem. Phys.}\
  }\textbf {\bibinfo {volume} {54}},\ \bibinfo {pages} {5237} (\bibinfo {year}
  {1971})}\BibitemShut {NoStop}%
\bibitem [{\citenamefont {Kremer}\ and\ \citenamefont
  {Grest}(1990)}]{kkremer-90}%
  \BibitemOpen
  \bibfield  {author} {\bibinfo {author} {\bibfnamefont {K.}~\bibnamefont
  {Kremer}}\ and\ \bibinfo {author} {\bibfnamefont {G.~S.}\ \bibnamefont
  {Grest}},\ }\href@noop {} {\bibfield  {journal} {\bibinfo  {journal} {J.
  Chem. Phys.}\ }\textbf {\bibinfo {volume} {92}},\ \bibinfo {pages} {5057}
  (\bibinfo {year} {1990})}\BibitemShut {NoStop}%
\bibitem [{\citenamefont {Plimpton}(1995)}]{splimpton-95}%
  \BibitemOpen
  \bibfield  {author} {\bibinfo {author} {\bibfnamefont {S.}~\bibnamefont
  {Plimpton}},\ }\href@noop {} {\bibfield  {journal} {\bibinfo  {journal} {J.
  Chem. Phys.}\ }\textbf {\bibinfo {volume} {117}},\ \bibinfo {pages} {1}
  (\bibinfo {year} {1995})}\BibitemShut {NoStop}%
\bibitem [{\citenamefont
  {Golestanian}(2009)}]{Ramin_PRL_anomalous_diffusion-09}%
  \BibitemOpen
  \bibfield  {author} {\bibinfo {author} {\bibfnamefont {R.}~\bibnamefont
  {Golestanian}},\ }\href@noop {} {\bibfield  {journal} {\bibinfo  {journal}
  {Phys. Rev. Lett.}\ }\textbf {\bibinfo {volume} {102}},\ \bibinfo {pages}
  {188305} (\bibinfo {year} {2009})}\BibitemShut {NoStop}%
\bibitem [{\citenamefont {Schol}\ \emph {et~al.}(2018)\citenamefont {Schol},
  \citenamefont {Jahanshahi}, \citenamefont {Ldov},\ and\ \citenamefont
  {L{\"o}wen}}]{inertia_natcom_Lowen-2018}%
  \BibitemOpen
  \bibfield  {author} {\bibinfo {author} {\bibfnamefont {C.}~\bibnamefont
  {Schol}}, \bibinfo {author} {\bibfnamefont {S.}~\bibnamefont {Jahanshahi}},
  \bibinfo {author} {\bibfnamefont {A.}~\bibnamefont {Ldov}}, \ and\ \bibinfo
  {author} {\bibfnamefont {H.}~\bibnamefont {L{\"o}wen}},\ }\href@noop {}
  {\bibfield  {journal} {\bibinfo  {journal} {Nat. Comm.}\ }\textbf {\bibinfo
  {volume} {9}},\ \bibinfo {pages} {5156} (\bibinfo {year} {2018})}\BibitemShut
  {NoStop}%
\bibitem [{\citenamefont {Doi}\ and\ \citenamefont
  {Edwards}(1986)}]{doi-edwards-86}%
  \BibitemOpen
  \bibfield  {author} {\bibinfo {author} {\bibfnamefont {M.}~\bibnamefont
  {Doi}}\ and\ \bibinfo {author} {\bibfnamefont {S.~F.}\ \bibnamefont
  {Edwards}},\ }\href@noop {} {\emph {\bibinfo {title} {The theory of polymer
  dynamics}}}\ (\bibinfo  {publisher} {Oxford University Press},\ \bibinfo
  {address} {Newyork},\ \bibinfo {year} {1986})\BibitemShut {NoStop}%
\bibitem [{\citenamefont {Landau}\ and\ \citenamefont
  {Lifschitz}(1963)}]{landau-elasticity}%
  \BibitemOpen
  \bibfield  {author} {\bibinfo {author} {\bibfnamefont {L.~D.}\ \bibnamefont
  {Landau}}\ and\ \bibinfo {author} {\bibfnamefont {E.~M.}\ \bibnamefont
  {Lifschitz}},\ }\href@noop {} {\emph {\bibinfo {title} {The theory of
  elasticity 2nd Edition}}}\ (\bibinfo  {publisher} {Pergamon Press, Oxford},\
  \bibinfo {address} {Newyork},\ \bibinfo {year} {1963})\BibitemShut {NoStop}%
\bibitem [{\citenamefont {Marchetti}\ \emph {et~al.}(2013)\citenamefont
  {Marchetti}, \citenamefont {Joanny}, \citenamefont {Ramaswamy}, \citenamefont
  {Liverpool}, \citenamefont {Prost}, \citenamefont {Rao},\ and\ \citenamefont
  {Simha}}]{marchetti-13}%
  \BibitemOpen
  \bibfield  {author} {\bibinfo {author} {\bibfnamefont {M.}~\bibnamefont
  {Marchetti}}, \bibinfo {author} {\bibfnamefont {J.}~\bibnamefont {Joanny}},
  \bibinfo {author} {\bibfnamefont {S.}~\bibnamefont {Ramaswamy}}, \bibinfo
  {author} {\bibfnamefont {T.}~\bibnamefont {Liverpool}}, \bibinfo {author}
  {\bibfnamefont {J.}~\bibnamefont {Prost}}, \bibinfo {author} {\bibfnamefont
  {M.}~\bibnamefont {Rao}}, \ and\ \bibinfo {author} {\bibfnamefont {R.~A.}\
  \bibnamefont {Simha}},\ }\href@noop {} {\bibfield  {journal} {\bibinfo
  {journal} {Rev. Mod. Phys.}\ }\textbf {\bibinfo {volume} {85}},\ \bibinfo
  {pages} {1143} (\bibinfo {year} {2013})}\BibitemShut {NoStop}%
\bibitem [{\citenamefont {Prathyusha}\ \emph {et~al.}(2018)\citenamefont
  {Prathyusha}, \citenamefont {Henkes},\ and\ \citenamefont
  {Sknepnek}}]{prathyusha_PRE-18}%
  \BibitemOpen
  \bibfield  {author} {\bibinfo {author} {\bibfnamefont {K.~R.}\ \bibnamefont
  {Prathyusha}}, \bibinfo {author} {\bibfnamefont {S.}~\bibnamefont {Henkes}},
  \ and\ \bibinfo {author} {\bibfnamefont {R.}~\bibnamefont {Sknepnek}},\
  }\href@noop {} {\bibfield  {journal} {\bibinfo  {journal} {Phys. Rev. E}\
  }\textbf {\bibinfo {volume} {97}},\ \bibinfo {pages} {022606} (\bibinfo
  {year} {2018})}\BibitemShut {NoStop}%
\bibitem [{\citenamefont {Duman}\ \emph {et~al.}(2018)\citenamefont {Duman},
  \citenamefont {Isele-Holder}, \citenamefont {Elgeti},\ and\ \citenamefont
  {Gompper}}]{Gompper-collective-polymer-18}%
  \BibitemOpen
  \bibfield  {author} {\bibinfo {author} {\bibfnamefont {{\"O}.}~\bibnamefont
  {Duman}}, \bibinfo {author} {\bibfnamefont {R.~E.}\ \bibnamefont
  {Isele-Holder}}, \bibinfo {author} {\bibfnamefont {J.}~\bibnamefont
  {Elgeti}}, \ and\ \bibinfo {author} {\bibfnamefont {G.}~\bibnamefont
  {Gompper}},\ }\href@noop {} {\bibfield  {journal} {\bibinfo  {journal} {Soft
  Matter}\ }\textbf {\bibinfo {volume} {14}},\ \bibinfo {pages} {4483}
  (\bibinfo {year} {2018})}\BibitemShut {NoStop}%
\bibitem [{\citenamefont {Arslanova}\ \emph {et~al.}(2021)\citenamefont
  {Arslanova}, \citenamefont {Dugyala}, \citenamefont {Reichel}, \citenamefont
  {Reddy}, \citenamefont {Fransaer},\ and\ \citenamefont
  {Clasen}}]{janussweeper-21}%
  \BibitemOpen
  \bibfield  {author} {\bibinfo {author} {\bibfnamefont {A.}~\bibnamefont
  {Arslanova}}, \bibinfo {author} {\bibfnamefont {V.~R.}\ \bibnamefont
  {Dugyala}}, \bibinfo {author} {\bibfnamefont {E.~K.}\ \bibnamefont
  {Reichel}}, \bibinfo {author} {\bibfnamefont {N.}~\bibnamefont {Reddy}},
  \bibinfo {author} {\bibfnamefont {J.}~\bibnamefont {Fransaer}}, \ and\
  \bibinfo {author} {\bibfnamefont {C.}~\bibnamefont {Clasen}},\ }\href@noop {}
  {\bibfield  {journal} {\bibinfo  {journal} {Soft Matter}\ }\textbf {\bibinfo
  {volume} {17}},\ \bibinfo {pages} {2369} (\bibinfo {year}
  {2021})}\BibitemShut {NoStop}%
\bibitem [{\citenamefont {Zhao}\ \emph {et~al.}(2019)\citenamefont {Zhao},
  \citenamefont {Xie}, \citenamefont {Liu}, \citenamefont {Liu}, \citenamefont
  {Song},\ and\ \citenamefont {Li}}]{motion-capture-bacterial-Linanoscale19}%
  \BibitemOpen
  \bibfield  {author} {\bibinfo {author} {\bibfnamefont {L.}~\bibnamefont
  {Zhao}}, \bibinfo {author} {\bibfnamefont {S.}~\bibnamefont {Xie}}, \bibinfo
  {author} {\bibfnamefont {Y.}~\bibnamefont {Liu}}, \bibinfo {author}
  {\bibfnamefont {Q.}~\bibnamefont {Liu}}, \bibinfo {author} {\bibfnamefont
  {X.}~\bibnamefont {Song}}, \ and\ \bibinfo {author} {\bibfnamefont
  {X.}~\bibnamefont {Li}},\ }\href@noop {} {\bibfield  {journal} {\bibinfo
  {journal} {Nanoscale}\ }\textbf {\bibinfo {volume} {11}},\ \bibinfo {pages}
  {17831} (\bibinfo {year} {2019})}\BibitemShut {NoStop}%
\bibitem [{\citenamefont {Zhao}\ \emph {et~al.}(2020)\citenamefont {Zhao},
  \citenamefont {Liu}, \citenamefont {Xie}, \citenamefont {Ran}, \citenamefont
  {Wei}, \citenamefont {Liu},\ and\ \citenamefont
  {Li}}]{fluroscene-tumour-Li-Chem.Eng.20}%
  \BibitemOpen
  \bibfield  {author} {\bibinfo {author} {\bibfnamefont {L.}~\bibnamefont
  {Zhao}}, \bibinfo {author} {\bibfnamefont {Y.}~\bibnamefont {Liu}}, \bibinfo
  {author} {\bibfnamefont {S.}~\bibnamefont {Xie}}, \bibinfo {author}
  {\bibfnamefont {P.}~\bibnamefont {Ran}}, \bibinfo {author} {\bibfnamefont
  {J.}~\bibnamefont {Wei}}, \bibinfo {author} {\bibfnamefont {Q.}~\bibnamefont
  {Liu}}, \ and\ \bibinfo {author} {\bibfnamefont {X.}~\bibnamefont {Li}},\
  }\href@noop {} {\bibfield  {journal} {\bibinfo  {journal} {Chem. Eng. J.}\
  }\textbf {\bibinfo {volume} {382}},\ \bibinfo {pages} {123041} (\bibinfo
  {year} {2020})}\BibitemShut {NoStop}%
\end{thebibliography}

%

\end{document}